\documentclass[amssymb,nobibnotes,superscriptaddress,longbibliography,notitlepage,twocolumn]{revtex4-1}

\usepackage{enumerate,amsmath}
\usepackage[dvipdfmx]{graphicx}
\usepackage{color} 

\usepackage{braket}
\usepackage{qcircuit}

\begin{document}

\title{Partially Fault-tolerant Quantum Computing Architecture with Error-corrected Clifford Gates and Space-time Efficient Analog Rotations}%

\author{Yutaro Akahoshi}
\email{akahoshi.yutaro@fujitsu.com}
\author{Kazunori Maruyama}
\author{Hirotaka Oshima}
\author{Shintaro Sato}
\affiliation{Quantum Laboratory, Fujitsu Research, Fujitsu Limited, \\4-1-1 Kawasaki, Kanagawa 211-8588, Japan}
\affiliation{Fujitsu Quantum Computing Joint Research Division, Center for Quantum Information and Quantum Biology, Osaka University, \\1-2 Machikaneyama, Toyonaka, Osaka, 565-8531, Japan}

\author{Keisuke Fujii}
\affiliation{Fujitsu Quantum Computing Joint Research Division, Center for Quantum Information and Quantum Biology, Osaka University, \\1-2 Machikaneyama, Toyonaka, Osaka, 565-8531, Japan}
\affiliation{Graduate School of Engineering Science, Osaka University, \\1-3 Machikaneyama, Toyonaka, Osaka, 560-8531, Japan}
\affiliation{Center for Quantum Information and Quantum Biology, Osaka University, 560-0043, Japan}
\affiliation{RIKEN Center for Quantum Computing (RQC), Wako Saitama 351-0198, Japan}

\begin{abstract}
Quantum computers are expected to bring drastic acceleration to several computing tasks against classical computers. 
Noisy intermediate-scale quantum (NISQ) devices, which have tens to hundreds of noisy physical qubits, are gradually becoming available, but it is still challenging to achieve useful quantum advantages in meaningful tasks at this moment. 
On the other hand, the full fault-tolerant quantum computing (FTQC) based on the quantum error correction (QEC) code remains far beyond realization due to its extremely large requirement of high-precision physical qubits. 
In this study, we propose a quantum computing architecture to close the gap between NISQ and FTQC. 
Our architecture is based on erroneous arbitrary rotation gates and error-corrected Clifford gates implemented by lattice surgery. 
We omit the typical distillation protocol to achieve direct analog rotations and small qubit requirements, and minimize the remnant errors of the rotations by a carefully-designed state injection protocol. 
Our estimation based on numerical simulations shows that, for early-FTQC devices that consist of $10^4$ physical qubits with physical error probability $p = 10^{-4}$, we can perform roughly $1.72 \times 10^7$ Clifford operations and $3.75 \times 10^4$ arbitrary rotations on 64 logical qubits. 
Such computations cannot be realized by the existing NISQ and FTQC architectures on the same device, as well as classical computers. 
We hope that our proposal and the corresponding development of quantum algorithms based on it bring new insights on realization of practical quantum computers in future. 
\end{abstract}

\maketitle
\section{Introduction} \label{sect:intro}
Quantum computers are expected to provide exponential speedup of computation in certain tasks including factoring~\cite{shor1999polynomial}, simulating quantum many-body systems~\cite{abrams1999quantum,aspuru2005simulated}, and linear algebraic operations~\cite{harrow2009quantum}. 
To realize such a quantum computer, 
the development of quantum computing devices in various physical systems has been actively carried out in recent years.
While fidelity and controllability are diverse, quantum devices with tens to hundreds of qubits have emerged and are referred to as noisy intermediate-scale quantum (NISQ) devices~\cite{preskill2018quantum}. 
Now we are entering an era of quantum computational supremacy~\cite{arute2019quantum,zhong2020quantum,wu2021strong,zhu2022quantum,madsen2022quantum}, where simulating the behavior of a quantum computer itself is becoming intractable for a classical computer. 
Unfortunately, it is still challenging to extract useful quantum advantages over the classical best approaches from NISQ devices for practically meaningful tasks.
Additionally, the classical simulation technology of a quantum computer using supercomputers has recently improved, and it has been reported that random quantum circuit sampling on Google's quantum computer in 2019 can be simulated in comparable time~\cite{liu2021closing}.

The problem with NISQ devices is that they cannot provide an ultimate solution to the noise issue. 
Qubits and the gate operations on them lose their quantum nature due to decoherence caused by undesirable interactions with the environment, which introduces errors into a quantum computer. 
Thus, useful tasks are difficult to perform reliably.
Furthermore, the variational quantum algorithm~\cite{cerezo2021variational} is based on the estimation of expectation values, and the number of measurements increases with the number of qubits. 
The accuracy becomes poor due to statistical errors as well as the effects of noise, and the optimization of the variational parameters becomes extremely difficult~\cite{cerezo2021variational}. 
This problem may be solved by improving the fidelity of quantum devices and by using techniques such as quantum noise mitigation~\cite{endo2021hybrid}, specifically designed for NISQ devices.
However, 
it is a nontrivial question whether quantum noise mitigation can solve the noise problem at a realistic sampling size for quantum computation of the 50- to 100-qubit level, which is difficult to simulate even with a classical computer~\cite{brandhofer2021special}.
The ultimate long-term solution is the realization of  fault-tolerant quantum computing (FTQC) by implementing quantum error correction (QEC).

Several experiments have demonstrated the viability of QEC~\cite{zhao2022realization,krinner2022realizing,Acharya2023}. 
Soon, error correction will allow us to store quantum information for longer than its physical coherence time, and to perform fault-tolerant logical gate operations.
However, non-Clifford gates, which are necessary ingredients for quantum speedup, 
are difficult to implement fault-tolerantly on the QEC codes such as the surface code. 
A special protocol called magic state distillation is employed
to implement a non-Clifford $T$ gate reliably~\cite{bravyi2012magic}. 
Furthermore, the arbitrary angle rotation gates on a logical qubit
require the huge number of $T$ gates when they are decomposed into Clifford and $T$ gates
via Solovay-Kitaev decomposition.
Together with the cost of magic state distillation and Clifford$+T$ gate decomposition, the realization of fully-fledged FTQC requires an overhead of hundreds of thousands to millions of qubits~\cite{gidney2021factor,yoshioka2022hunting,reiher2017elucidating,goings2022reliably}.

In terms of the number of qubits for algorithm viability, a large gap will exist between the NISQ and FTQC eras; the number of qubits that is needed for a meaningful quantum computation differs by several orders of magnitude. 
While experimental breakthroughs are expected to emerge to integrate 1 million qubits in the long term, 
in the meantime it is also necessary to establish a theoretical framework that meaningfully exploits early FTQC with $10^3$--$10^4$ qubits. 

In this study, we propose a framework to hybridize NISQ and FTQC to close the gap between them and provide evidence that a quantum computer of ten thousand qubits has great potential to exhibit quantum advantages in meaningful tasks. 
In this direction, quantum noise mitigation designed for NISQ devices has been applied for FTQC to reduce the required number of physical qubits, while magic state distillation still requires huge number of physical qubits for quantum advantage~\cite{suzuki2022quantum,piveteau2021error}. 
Here, we integrate the NISQ and FTQC approaches at a deeper level. More concretely, in our approach, continuous rotational gates are not protected by QEC but are executed by injecting ancilla states without magic state distillation. 
This allows us to perform a rotation gate by an arbitrary angle directly. As a drawback, the injection of the ancilla states on a QEC code suffers from unavoidable errors. 
In our proposal, we carefully design an injection circuit so that most errors during the injection are detected and/or corrected so that the resultant special ancilla states have a minimum error. 
Furthermore, for the Clifford gates, such as CNOT, $H$, and $S$, we use the rotated planar surface code as usual, and hence errors are corrected in a scalable way. 
This allows us an almost error-free implementation of logical Clifford operations against the rotation gates. 

Combining error-corrected Clifford gates and reasonably clean analog rotations, our resource estimation shows that $3.75 \times 10^4$ arbitrary rotation gates and $1.72 \times 10^7$ Clifford gates on 64 logical qubits are reliably executed using $10^4$ physical qubits when the physical error probability is $10^{-4}$. 
Such computations cannot be simulated on classical computers, and even the existing NISQ and FTQC architectures on the same device cannot realize this amount of computational power. 
Our architecture can be applied to useful tasks such as the quantum many-body simulation and the quantum approximation optimization algorithm (QAOA) thanks to the fast implementation of the analog rotations.
The proposed Space-Time efficient Analog Rotation quantum computing architecture (hereinafter referred to as STAR architecture) provides a new framework for the use of quantum computers that fills the gap between the NISQ and FTQC eras. 

\section{An overview of the STAR architecture}
Before delving into a comprehensive description, we provide an overview of the STAR architecture in this section.

Now that we are in the NISQ era and the number of qubits is increasing to hundreds. 
However, it will be extremely difficult to fully exploit the computational power of NISQ devices with hundreds to thousands of qubits,
because NISQ devices suffer from errors in both Clifford and non-Clifford operations.
The number of gates increases due to the swap operations at the stage of compiling a quantum algorithm to be executable on actual quantum computing devices with limited qubit connectivity. 
Moreover, in applications to quantum chemistry, fermionic rotations, such as the unitary coupled cluster ansatz, require the entangling gates to generate multi-Pauli rotations. 
Most of the gates employed there are Clifford gates such as CNOT which do not make classical simulation difficult from the viewpoint of the Gottesman-Knill theorem, while they result in the accumulation of errors. 
As a result, the total number of non-Clifford gates that can be executed is rather limited.

Quantum error correction (QEC) is a method for entangling multiple qubits and encoding quantum information in a special subspace to protect it from noise~\cite{fujii2015quantum}. 
Unlike a classical bit, a qubit, which takes a superposed state through continuous complex probability amplitudes, suffers from continuous analog noise. 
The orthogonal subspace structure introduced by a QEC code can collapse such an analog noise into digitalized Pauli $X$, $Y$, and $Z$ errors, which are corrected appropriately. 
However, the orthogonal subspace structure also makes operations of the encoded degrees of freedom difficult~\cite{eastin2009restrictions}. 
Particularly, a fault-tolerant implementation of the $T$ gate, which is a non-Clifford gate and an essential ingredient for universal quantum computation, on an encoded degree of freedom is highly nontrivial. 
Most QEC codes do not support fault-tolerance for the $T$ gate in a native way.
A special protocol, called magic state distillation~\cite{bravyi2012magic}, is necessary to purify noisy magic states and execute $T$ gate via gate teleportation~\cite{zhou2000methodology}. 
Furthermore, since the $T$ gate is an $\pi/8$ rotation gate around the $z$-axis, an arbitrary rotational gate has been complied to  the sequence of Clifford gates and $T$ gates by using the Solovay-Kitaev algorithm. 
The state-of-the-art optimal Clifford$+T$ decomposition~\cite{ross2016optimal} still requires several tens of $T$ gates to achieve the $10^{-4}$ accuracy of an arbitrary single-qubit rotational gate, even if this accuracy can be achievable by physical single qubit rotation. 

\begin{figure*}[tbp]
  \centering 
  \includegraphics[width=130mm, clip]{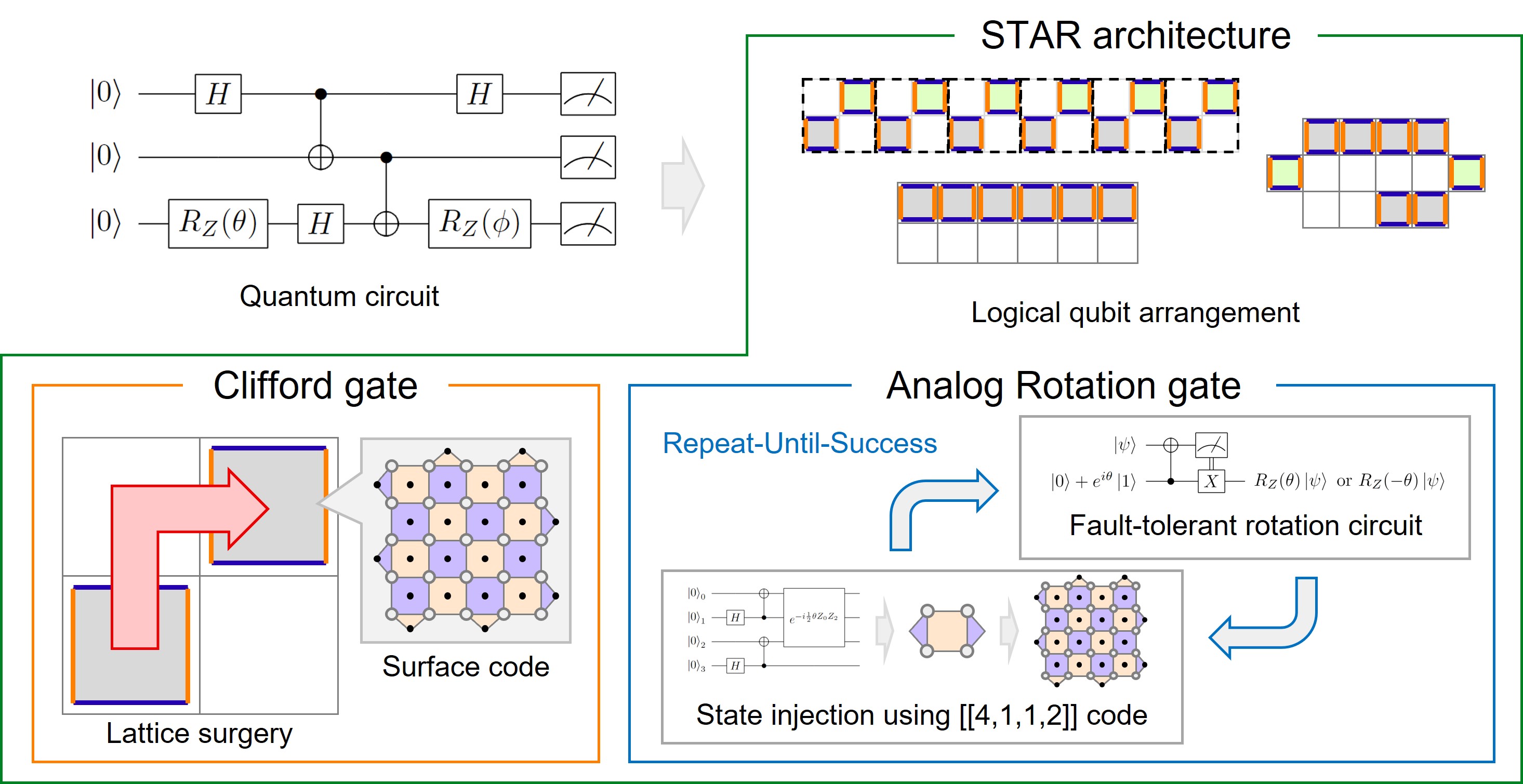}
  \caption{Overview of the STAR architecture proposed in this work. 
  The STAR architecture performs universal quantum computation using the error-corrected Clifford gate and analog rotation gate. 
  Clifford gates are implemented by the lattice surgery based on the rotated planar surface code. 
  Analog rotation is directly performed avoiding the magic state distillation, and 
  a special ancilla state needed for the rotation is cleanly injected into the rotated planar surface code by error-detection and post-selection. 
  Combined with an appropriate logical qubit arrangement, the STAR architecture fully exploits the computational power of early FTQC devices.  
  }
  \label{fig:overview}
\end{figure*}

If a quantum computer can execute all Clifford operations ideally and errors are introduced only in analog non-Clifford operations, more advanced quantum algorithms can be executed even in the era of early FTQC.
Our approach is to construct such architecture by successfully combining fault-tolerant error correction in FTQC and analog operations in NISQ. 
An overview of the STAR architecture is summarized in Fig.~\ref{fig:overview}. 
The key points are as follows:
\begin{enumerate}[(i)]
\item Fault-tolerant Clifford gates with QEC.
\item Analog rotation gates with reasonably clean ancilla state injection.
\end{enumerate}
Regarding (i), since the Clifford gates are protected by QEC 
the connectivity of physical qubits and Clifford transforms for gates such as many-body Pauli rotations
are not limiting factors to design reliable quantum computing.
We use the rotated planar surface code~\cite{Horsman_2012} as a logical qubit 
and employ lattice surgery~\cite{Litinski2019gameofsurfacecodes} to implement the fault-tolerant logical Clifford gates,
as will be explained in Sec.~\ref{sec:ftcg}.

On the other hand, by virtue of (ii),
we can avoid the magic state distillation, which is the costly part of FTQC.
This also successfully reduces the computational cost in a double sense in that it eliminates the need for decomposition into $T$ gates when performing continuous rotation gates.
As explained in Sec.~\ref{sec:arg},
we carefully design a quantum circuit to inject a special ancilla state into the planar surface code
with error detection and post-selection. Then the reasonably clean ancilla states are used to implement 
analog rotation gates via gate teleportation, where the byproduct is treated 
by a repeat-until-success (RUS) approach.
In Sec.~\ref{sec:lqa}, we provide typical logical qubit arrangements in the STAR architecture. 
As shown in Sec.~\ref{sect:resource_estimation}, according to our numerical simulations, the STAR architecture surpasses the existing NISQ and FTQC architectures on the early-FTQC device as well as the classical computers. 
We also discuss the possible applications of our architecture there. 
Sec.~\ref{sec:conclusion} concludes this study and discusses future directions. 


\section{Fault-tolerant Clifford gates}
\label{sec:ftcg}
To make the discussion self-contained, we will start by reviewing the existing approaches
for encoding quantum information into the rotated planar surface code and protecting Clifford gates on them.

\subsection{Rotated planar surface code} 
\begin{figure}[tbp]
  \centering
  \includegraphics[width=40mm, clip]{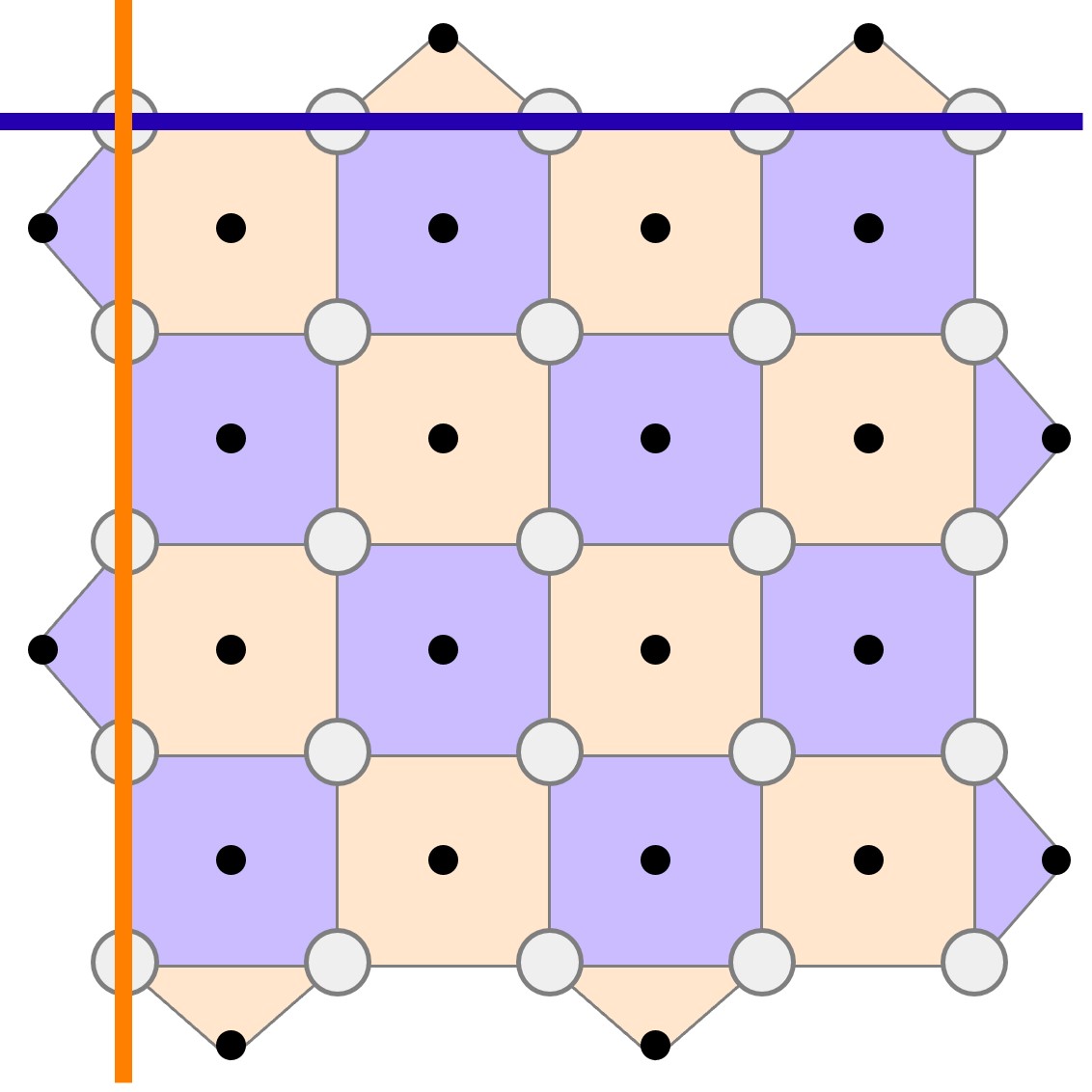}
  \caption{Definition of the rotated planar surface code with the code distance $d = 5$. 
  White and black circles represent physical qubits constructing a logical state and measurement qubits, respectively. 
  Stabilizer operators, which define the logical state, are shown as orange and blue colored surfaces. 
  Representative logical operators are given in solid lines on boundaries.}
  \label{fig:surface_code_def}
\end{figure}

The rotated planar surface code is a QEC code that has good features suitable for the early-FTQC devices: 
a relatively high threshold value against other QEC codes and a small requirement for the number of physical qubits\cite{Horsman_2012,PhysRevA.89.022321}. 
We summarize its definition in Fig.~\ref{fig:surface_code_def}. 
Physical qubits constructing the rotated planar surface code are arranged on the vertices of a two-dimensional lattice (white circles in Fig.~\ref{fig:surface_code_def}). 
$X$ ($Z$) stabilizer operators are defined on the faces of the lattice (orange (blue) faces in Fig.~\ref{fig:surface_code_def}),  
and the logical state is defined as the simultaneous eigenstate of those stabilizer operators with eigenvalues of $+1$.
A single surface has two types of boundaries, namely the $X$- and $Z$-boundary, 
along which the logical $X$ and $Z$ operators are defined as chains of physical $X$ and $Z$ operators 
(orange and blue lines on the boundaries in Fig.~\ref{fig:surface_code_def}). 
The code distance $d$ is equal to the length of a side of the lattice. 
In the following discussion, we call a surface that carries a single logical state a ``logical patch" or simply a ``patch". 

\begin{figure}[tbp]
  \centering
  \begin{tabular}{cc}
    \includegraphics[width=22mm, clip]{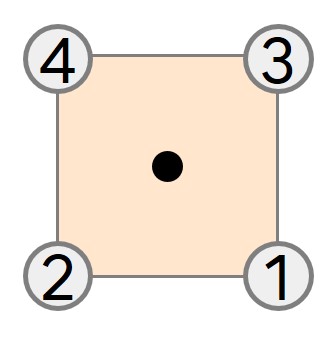} &
    \includegraphics[width=40mm, clip]{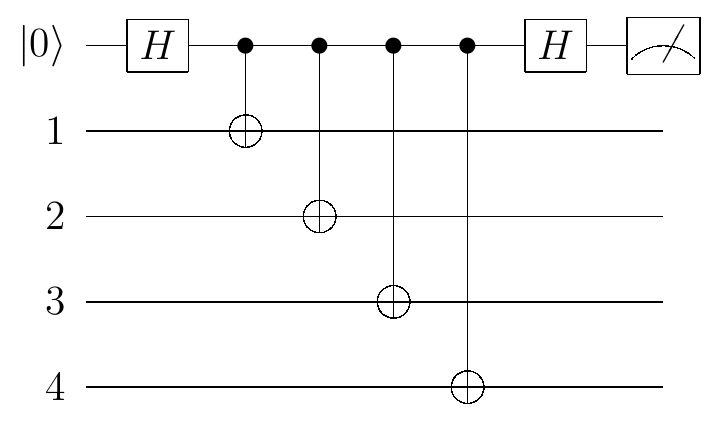} 
    \\ 
    \includegraphics[width=22mm, clip]{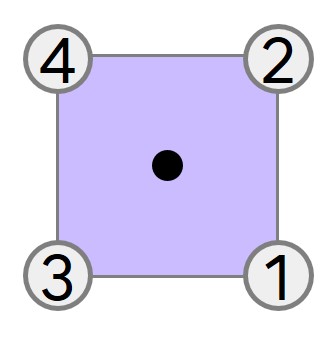} & 
    \includegraphics[width=40mm, clip]{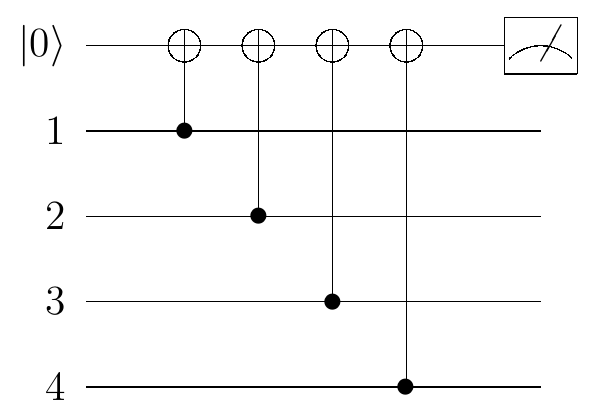}   
  \end{tabular}
  \caption{Syndrome measurement circuits. 
  (Upper row) $X$ syndrome measurement circuit. 
  The order of CNOT operations is represented by circled numbers on the left hand side. 
  (Lower row) The same figure for the $Z$ syndrome measurement circuit.}
  \label{fig:meas_circ}
\end{figure}

To perform the error correction, one measures the eigenvalues of the stabilizers 
using measurement qubits arranged on the faces of the lattice (black circles in Fig.~\ref{fig:surface_code_def}). 
Measured eigenvalues are called ``syndromes" and are utilized to infer a possible error pattern. 
The syndrome measurement circuits that we employ in this study are shown in Fig.~\ref{fig:meas_circ}. 
Using this circuit, we can measure simultaneously all syndromes in eight fundamental operation steps. 
The order of CNOT operations between physical and measurement qubits in Fig.~\ref{fig:meas_circ} is important 
for preserving commutation relations between stabilizer operators. 
In this study, we employ the order proposed in Ref.~\cite{Litinski2018latticesurgery} to prevent hook errors along the logical operators. 

Errors occurred in the logical qubit are inferred by observed error syndromes, where the eigenvalue of the stabilizer is flipped from $+1$ to $-1$. 
In the surface code, we can infer the most likely error pattern as follows. 
First, we construct a decoder graph, in which syndromes and error events are represented by nodes and edges, respectively. 
Paths that connect error syndromes in the graph provide candidates for the actual error pattern, 
and their length is related to the number of errors. 
Therefore, we can adopt the shortest path among these candidates as the most likely error pattern by assuming that the errors occur independently. 
The shortest path connecting the error syndrome is determined by a certain matching algorithm, e.g., the Edmonds' minimum-weight perfect matching (MWPM) algorithm~\cite{Edmonds1965}.
In practice, measured syndromes are also unreliable due to measurement errors; thus 
the syndrome measurement is repeated $d$ times and their differences in time axis are calculated by taking the XOR operation of the temporally neighboring two syndromes (in the following, we call these differences ``syndromes" unless otherwise stated). 
Then, we can construct a spatiotemporal decoder graph from $d$ sets of syndromes and infer the most likely error chains including the measurement errors by the MWPM algorithm. 
In addition to the MWPM algorithm, several other ways to perform this error inference have been proposed, such as  
the Union-Find algorithm~\cite{Delfosse2021almostlineartime}, the renormalization group decoder~\cite{PhysRevLett.104.050504}, and the Ising model-based approach~\cite{PhysRevX.4.041039,PhysRevResearch.4.043086}. 
In this study, we employ the MWPM algorithm to benchmark the performance of the logical patch. 

\subsection{Clifford gates by lattice surgery} \label{sect:latticesurgery}
In principle, logical Clifford gates like CNOT gate and Hadamard gate can be transversally perfomed in the planar rotated surface code~\cite{Horsman_2012}. 
In practice, however, the transversal CNOT gate is difficult to realize for some devices in which the connectivity between physical qubits is restricted. 
A clever way to implement Clifford gates in this situation is the lattice surgery, which consists of two-patch merging, splitting and patch deformation~\cite{Horsman_2012,Litinski2019gameofsurfacecodes}. 
The Clifford gates implemented in the STAR architecture rely on this technique. 
Here, we discuss typical examples to implement the logical CNOT gate and the logaical Hadamard gate, based on the fundamental operations introduced in Ref.~\cite{Litinski2019gameofsurfacecodes}. 

\begin{figure}[tbp]
  \centering
  \includegraphics[width=70mm, clip]{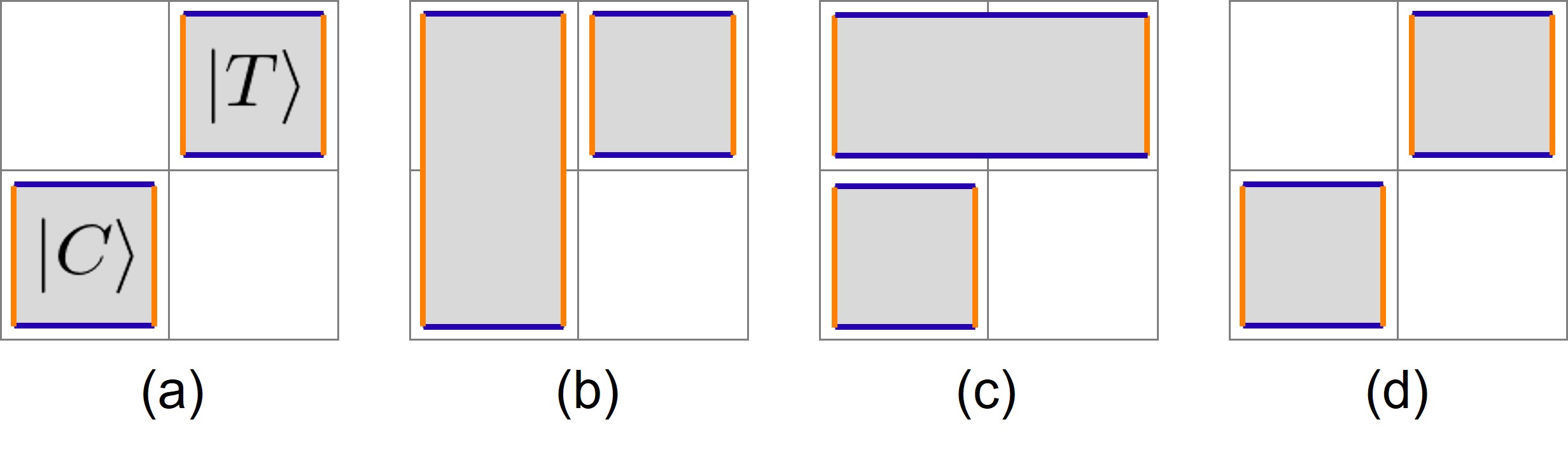}
  \caption{Logical CNOT operation by lattice surgery. Blue (orange) lines indicate $Z$- ($X$-) boundaries. 
  (a) Initial configuration of two logical patches. 
  (b) Expand the control patch along the $X$-boundary. 
  (c) Split the control patch along the $Z$-boundary and merge one of them with the target patch along the $X$-boundary. These two operations can be performed simultaneously. 
  (d) Contract the target patch along the $Z$-boundary.}
  \label{fig:surgery_cnot}
\end{figure}
\begin{figure}[tbp]
  \centering
  \includegraphics[width=70mm, clip]{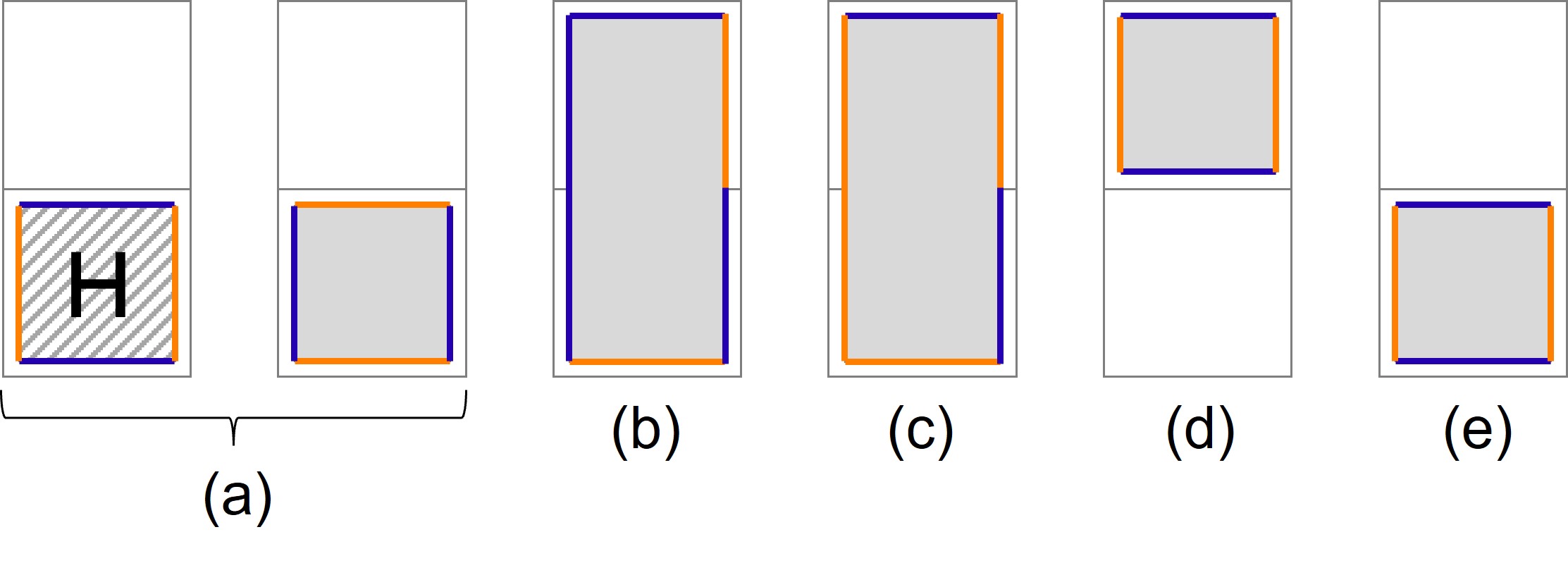}
  \caption{Logical Hadamard operation with a boundary rotation by lattice surgery.
  (a) After acting the transversal Hadamard gate on a certain patch, its orientation rotates $90^{\circ}$. 
  (b) To fix its orientation, expand the patch first. 
  (c) Modify its boundary
  (d) Contract the patch. At this moment its orientation is fixed correctly. 
  (e) Move the patch to the original position. This operation can be achieved by combining patch expansion and contraction. }
  \label{fig:surgery_h}
\end{figure}
A standard logical CNOT operation using the lattice surgery is achieved by merging and splitting a control logical qubit $\ket C$ and a target logical qubit $\ket T$. 
Figure~\ref{fig:surgery_cnot} shows a sequence of the lattice surgery operations for performing the logical CNOT operation. 
The logical state $\ket C$ and $\ket T$ are placed as Fig.~\ref{fig:surgery_cnot} (a). 
Then, the following lattice surgery operations are performed: 
(b) expand $\ket C$ along the $X$-boundary, 
(c) split $\ket C$ into two patches by the $Z$-boundary and merge one of them with $\ket T$ along the $X$-boundary, and
(d) contract $\ket T$ along the $Z$-boundary. 
Operations (b) and (c) need $d$ rounds of the syndrome measurement to determine syndrome values, and a total of $2d$ rounds for the logical CNOT operation. 
If the measured $X_L \otimes X_L$ eigenvalue, which is a product of the eigenvalues of the $X$ stabilizers newly introduced in the $X$-boundary merging in (c),  is equal to $-1$, a byproduct operator $Z_L$ subsequently acts on $\ket C$. 

A logical Hadamard gate is simply achieved by transversally acting a physical Hadamard gate on all data qubits. 
An important obstacle is that the logical qubit patch after the operation 
rotates $90^{\circ}$
from the original orientation [Fig.~\ref{fig:surgery_h}(a)]. 
This rotated orientation can be corrected by the lattice surgery operations. 
A typical sequence of the operation is shown in Fig.~\ref{fig:surgery_h}: 
(b) expand a patch, (c) deform the patch boundary, (d) contract the patch, and (e) move the patch to the original position. 
In this example, operations (b), (c) and (e) need $d$ rounds of the syndrome measurement; thus $3d$ rounds are required in total. 

\begin{figure}[bp]
  \centering
  \includegraphics[width=30mm, clip]{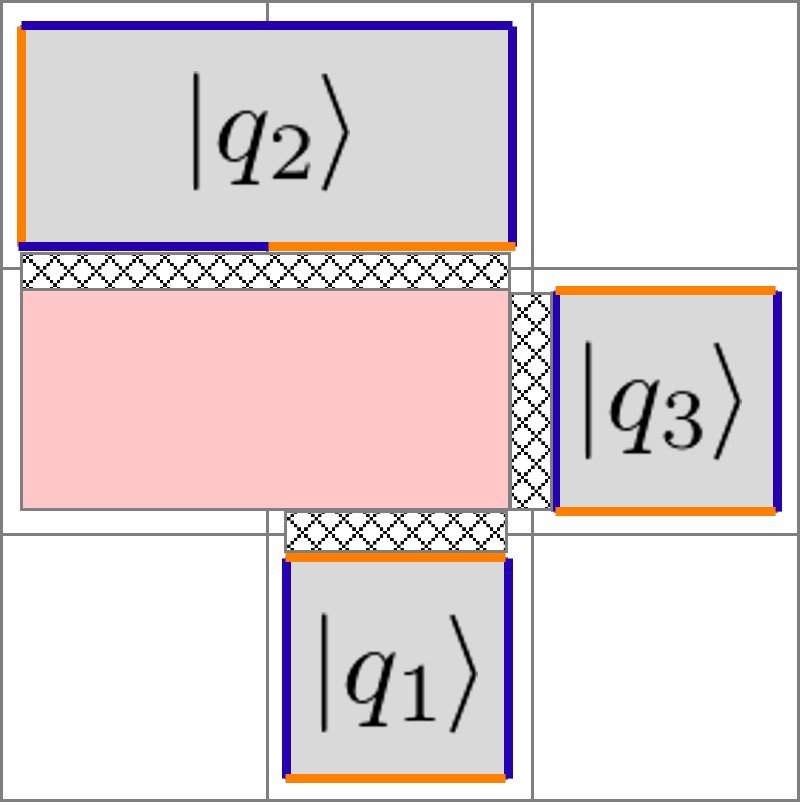}
  \caption{Example of the multi-Pauli measurement $X_{L,1} \otimes Y_{L,2} \otimes Z_{L,3}$. }
  \label{fig:multip_meas}
\end{figure}

In Ref.~\cite{Litinski2019gameofsurfacecodes}, the author proposes another way to perform fault-tolerant quantum computation, 
in which the Clifford gates in quantum circuits are moved to the end of the circuits and absorbed into measurements. 
The modified circuits contain multi-Pauli measurements and multi-Pauli $\pi/8$ rotations, which can also be performed by the lattice surgery. 
A typical example of measuring a multi-Pauli operator $X_{L,1} \otimes Y_{L,2} \otimes Z_{L,3}$ is given in Fig.~\ref{fig:multip_meas}. 
Physical qubits in an ancilla region are first initialized to $\ket +$ (a red region in Fig.~\ref{fig:multip_meas}), and 
then the stabilizer operators in the entire region are measured (including the hatched area in Fig.~\ref{fig:multip_meas}).
The product of the eigenvalues of stabilizer operators whose eigenvalues are not determined by the initialization gives the measurement result of $X_{L,1} \otimes Y_{L,2} \otimes Z_{L,3}$. 
This operation is performed by $d$ rounds of syndrome measurement. 

The Clifford operations discussed in this section are closely related to the arrangement of logical qubits, 
which plays an important role in performing quantum computations with small overheads. 
We discuss typical examples of the arrangement later in Sec.~\ref{sec:lqa}. 


\section{Space-time efficient analog rotation gate}
\label{sec:arg}
In this section, we discuss how to implement analog rotation gates within a reasonable accuracy, that is a core technology of the STAR architecture. 
We directly perform analog rotation gates without the lengthy Solovay-Kitaev decomposition and avoid the costly magic state distillation. 
This approach is advantageous in terms of a physical qubit requirement and execution time. 
However, a major challenge is that the logical error rate of the analog rotation becomes relatively large at $O(p)$. 
To minimize the logical error rate, in our proposal, we carefully design a state injection protocol needed to generate a special ancilla state for the rotation. 
The remaining logical error of the analog rotation becomes a simple phase-flip error and can be further mitigated by the probabilistic error cancellation when the physical error probability is sufficiently small. 

\subsection{Repeat-until-success implementation of analog rotation gate} 
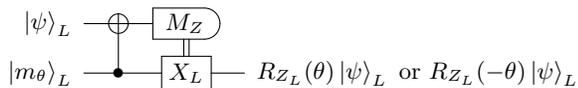
\begin{figure}[tbp]
\hspace{-40mm}
  \mbox{
  \Qcircuit @C=1em @R=.7em {
    \lstick{\ket{\psi}_L}                      & \targ     & \measureD{M_{Z}}  &  \\
    \lstick{\ket{m_\theta}_L} & \ctrl{-1} & \gate{X_L} \cwx  & \rstick{R_{Z_L}(\theta) \ket{\psi}_L \ {\rm or}\ R_{Z_L}(-\theta) \ket{\psi}_L } \qw 
    }
  }
  \caption{Quantum circuit for the analog $Z$ rotation gate. $M_Z$ is a destructive $Z_L$ measurement on a logical patch.}
  \label{fig:GT_circ}
\end{figure}

In the typical Clifford + $T$ gate decomposition in FTQC, the $T$ gate is implemented by the gate teleportation circuit with the magic state. 
Furthermore, to achieve an analog rotation gate with a sufficient accuracy, approximately 100 $T$ gates are required via the Solovay-Kitaev decomposition~\cite{ross2016optimal}. 
In contrast, the STAR architecture directly implements the analog rotation gate by using a special ancilla state $\ket{m_\theta} \equiv R_Z(\theta) \ket + = \frac{1}{\sqrt{2}}(e^{-i \theta / 2} \ket 0 + e^{+i \theta / 2} \ket 1)$, 
where the angle $\theta$ can be chosen arbitrarily. 
The circuit for implementing the analog rotation is shown in Fig.~\ref{fig:GT_circ}.
Since we allow arbitrary rotation angles, this implementation is not deterministic: 
An output state is a correctly rotated state $R_Z(\theta) \ket \psi$ if the measurement result in the circuit is $+1$; 
otherwise the output is an inversely rotated state, $R_Z(-\theta) \ket \psi$. 
Both outputs evenly occur. 
If the inversely rotated state is obtained, we apply a rotation gate with an angle $2 \theta$ on the output state 
to correct its angle. 
This correction is repeated until obtaining $R_Z(\theta) \ket \psi$ ("Repeat Until Success" or RUS). 
An average RUS step number to succeed is given as  
\begin{equation}
  1 \times \frac{1}{2} + 2 \times \frac{1}{4} + 3 \times \frac{1}{8} + \cdots = \sum_{i=1}^{\infty} \frac{n}{2^n} = 2.
\end{equation}
The Clifford gates in Fig.~\ref{fig:GT_circ} are performed by the lattice surgery. 

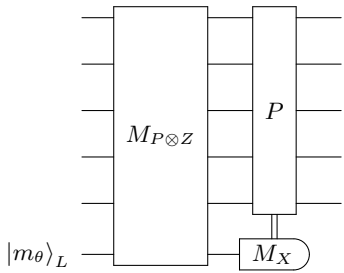
\begin{figure}[tbp]
  \centering
  \mbox
  {
  \Qcircuit @C=1.3em @R=1.0em {
    & \multigate{5}{M_{P \otimes Z}} & \multigate{4}{P} & \qw\\
    & \ghost{M_{P \otimes Z}} & \ghost{P} & \qw\\
    & \ghost{M_{P \otimes Z}} & \ghost{P} & \qw\\
    & \ghost{M_{P \otimes Z}} & \ghost{P} & \qw\\
    & \ghost{M_{P \otimes Z}} & \ghost{P} & \qw\\
    \lstick{\ket{m_\theta}_L} & \ghost{M_{P \otimes Z}} & \measureD{M_X} \cwx & 
  }
  }
  \caption{Quantum circuit for the analog multi-Pauli rotation gate. }
  \label{fig:mprot_circ}
\end{figure}
In the computational scheme using the multi-Pauli measurement and the multi-Pauli $\pi/8$ rotation gate as mentioned in the previous section, 
the $\pi/8$ rotation gate needs to be extended to an arbitrary angle multi-Pauli $P$ rotation gate (e.g. $P = X \otimes Y \otimes Z$). 
Such multi-Pauli rotation gates can be realized by performing a multi-Pauli $P \otimes Z$ measurement with target logical qubits and the ancilla state $\ket{m_{\theta}}$~\cite{Litinski2019gameofsurfacecodes}, 
as in the circuit shown in Fig.~\ref{fig:mprot_circ}. 
If the measurement result of $P \otimes Z$ is $+1$, the rotation succeeds, otherwise we must apply $R_P(2\theta)$ to the output state to correct its angle. 
An $X$ measurement in the circuit checks whether the output state has a byproduct operator $P$.
Since the byproduct $P$ commutes with $R_P(\theta)$ and satisfies $P^2 = I$, it is sufficient to cancel it after completing the entire RUS protocol if a product of the all $X$ measurement values is equal to $-1$.

\subsection{Low-error state injection protocol} \label{sect:state_prep} 
As discussed above, an analog rotation is implemented by circuits comprising error-corrected Clifford gates. 
Therefore, the accuracy of the analog rotation is dominated by the state injection protocol of the special ancilla state $\ket{m_{\theta}}$. 
In this section, we discuss a low-error state injection protocol based on the post-selection. 
This post-selection procedure is independent of the data logical patches involved in the main calculation and is scalable to the overall size of the calculation.

The first step of our injection protocol is to generate the ancilla state encoded in the $[[4,1,1,2]]$ subsystem code~\cite{PhysRevA.73.012340}. 
This code is defined by four physical qubits (we index them by subscripts $0$--$3$ in the following discussion) with two stabilizer operators
\begin{equation} \label{eq:422stabs}
  S_X = X_0 X_1 X_2 X_3, \quad S_Z = Z_0 Z_1 Z_2 Z_3.
\end{equation}
The $+1$ eigenstate of these stabilizers defines a logical qubit with logical operators 
\begin{equation} \label{eq:422logops_1}
  L_X = X_0 X_1, \quad L_Z = Z_0 Z_2,
\end{equation}
and gauge operators 
\begin{equation} \label{eq:422logops_2}
  G_X = X_0 X_2, \quad G_Z = Z_0 Z_1. 
\end{equation}
The code distance is two, so it can detect a single error. 

\begin{figure}[tbp]
  \centering
  \mbox
  {
  \Qcircuit @C=1.3em @R=1.0em {
    \lstick{\ket{0}_0} &\qw      & \targ     & \multigate{2}{R_{Z_0 Z_2}(\theta)} & \qw \\
    \lstick{\ket{0}_1} &\gate{H} & \ctrl{-1} & \ghost{R_{Z_0 Z_2}(\theta)} & \qw \\
    \lstick{\ket{0}_2} &\qw      & \targ     & \ghost{R_{Z_0 Z_2}(\theta)} & \qw \\
    \lstick{\ket{0}_3} &\gate{H} & \ctrl{-1} & \qw                         & \qw
  }
  }
  \caption{Quantum circuit for the injection of the ancilla state encoded in the $[[4,1,1,2]]$ subsystem code. }
  \label{fig:422prep_circ}
\end{figure}
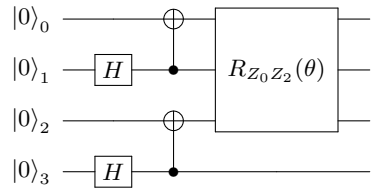
A circuit for preparing the ancilla state $\ket{m_\theta}_L$ encoded in the $[[4,1,1,2]]$ subsystem code is shown in Fig.~\ref{fig:422prep_circ}.
Hadamard and CNOT operations encode the input state into ${\ket +}_L$ state first, and 
then the logical rotation gate $R_{Z_0 Z_2}(\theta) \equiv e^{-i \theta / 2 (Z_0  Z_2)}$ acts on ${\ket +}_L$. 
The output state is therefore $\ket{m_\theta}_L$ up to an irrelevant overall factor. 
We assume that the gate $R_{Z_0 Z_2}(\theta) \equiv e^{-i \theta / 2 (Z_0  Z_2)}$ can be directly performed here. 

The generated ancilla state may suffer from errors in practice; thus, we measure syndromes of the $[[4,1,1,2]]$ subsystem code. 
Naively, we must measure the weight-4 stabilizer operator defined in Eq.(\ref{eq:422stabs}). 
Because of the gauge degrees of freedom (DOF), however, 
we can measure them as products of weight-2 gauge operators, whose measurements do not collapse the logical qubit. 
This property reduces noise in the ancilla state 
since it avoids critical weight-2 hook errors propagating from the measurement qubits and reduces the depth of the measurement circuit. 
As shown in Fig.~\ref{fig:422code_meascirc1}, 
we assume four measurement qubits that interact with the two nearest physical qubits (black circles labeled as M0--M3).  
This arrangement can be smoothly embedded in the rotated surface code as discussed later. 
The measurement circuit based on this arrangement is shown in Fig.~\ref{fig:422code_meascirc2}. 
To detect measurement errors, the measurement circuit is repeated twice, and we discard the prepared state 
if the measured syndromes satisfy one of the following conditions (post-selection): 
(i) one (or both) of the syndromes is equal to $-1$, or
(ii) one (or both) of the bare (the XOR operation is not performed) syndromes of the first round takes $-1$, although the syndromes are equal to $+1$. 
The state injection circuit is repeated until passing this post-selection.  
\begin{figure}[tbp]
  \centering
  \includegraphics[width=40mm, clip]{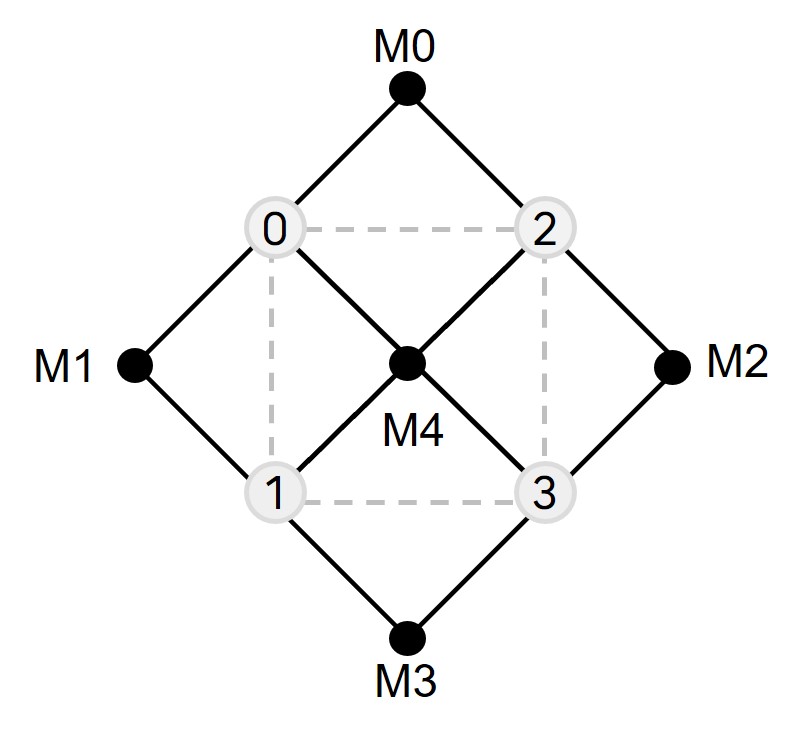}
  \caption{Arrangement of physical and measurement qubits encoded in the $[[4,1,1,2]]$ subsystem code. 
  White circles labeled as 0--3 and black circles labeled as M0--M3 are physical and measurement qubits, respectively. 
  The solid line shows the connectivity between measurement qubits and physical qubits.}
  \label{fig:422code_meascirc1}
\end{figure}
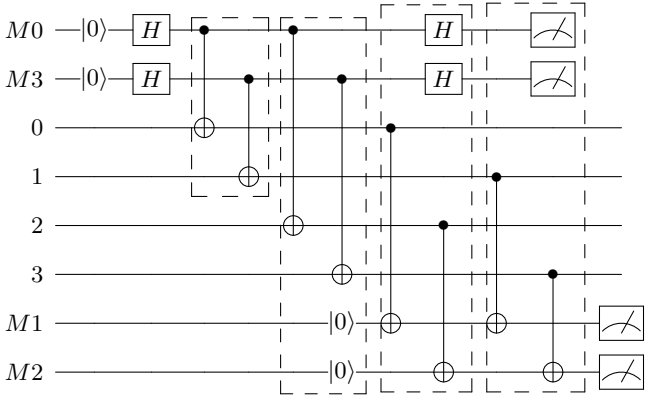
\begin{figure}[tbp]
  \centering
  \mbox{
      \Qcircuit @C=1em @R=.6em @!R {
        \lstick{M0}& \push{\ket 0} \qw & \gate{H} & \ctrl{2} & \qw & \ctrl{4} & \qw & \qw & \gate{H} & \qw & \meter \\
        \lstick{M3}& \push{\ket 0} \qw & \gate{H} & \qw & \ctrl{2} & \qw & \ctrl{4} & \qw & \gate{H} & \qw & \meter \\ 
        \lstick{0} & \qw & \qw & \targ & \qw & \qw & \qw & \ctrl{4} & \qw & \qw & \qw & \qw \\
        \lstick{1} & \qw & \qw & \qw & \targ & \qw & \qw & \qw & \qw & \ctrl{3} & \qw & \qw \\
        \lstick{2} & \qw & \qw & \qw & \qw & \targ & \qw & \qw & \ctrl{3} & \qw & \qw & \qw \\
        \lstick{3} & \qw & \qw & \qw & \qw & \qw & \targ & \qw & \qw & \qw & \ctrl{2} & \qw \\
        \lstick{M1}& \qw & \qw & \qw & \qw & \qw & \push{\ket 0} \qw & \targ & \qw & \targ & \qw & \meter \\
        \lstick{M2}& \qw & \qw & \qw & \qw & \qw & \push{\ket 0} \qw & \qw & \targ & \qw & \targ & \meter
        \gategroup{1}{4}{4}{5}{.8em}{--}
        \gategroup{1}{6}{8}{7}{.8em}{--}
        \gategroup{1}{8}{8}{9}{.8em}{--}
        \gategroup{1}{10}{8}{11}{.8em}{--}
      }
    }
    \caption{Syndrome measurement circuit of the $[[4,1,1,2]]$ subsystem code using the gauge operators. 
    The labels of physical and measurement qubits are the same in Fig.~\ref{fig:422code_meascirc1}. Meter symbols indicate $Z$ measurements. 
    Gates grouped by dashed lines are simultaneously implemented.}
    \label{fig:422code_meascirc2}
\end{figure}

\begin{figure}[b]
  \centering
  \includegraphics[width=80mm, clip]{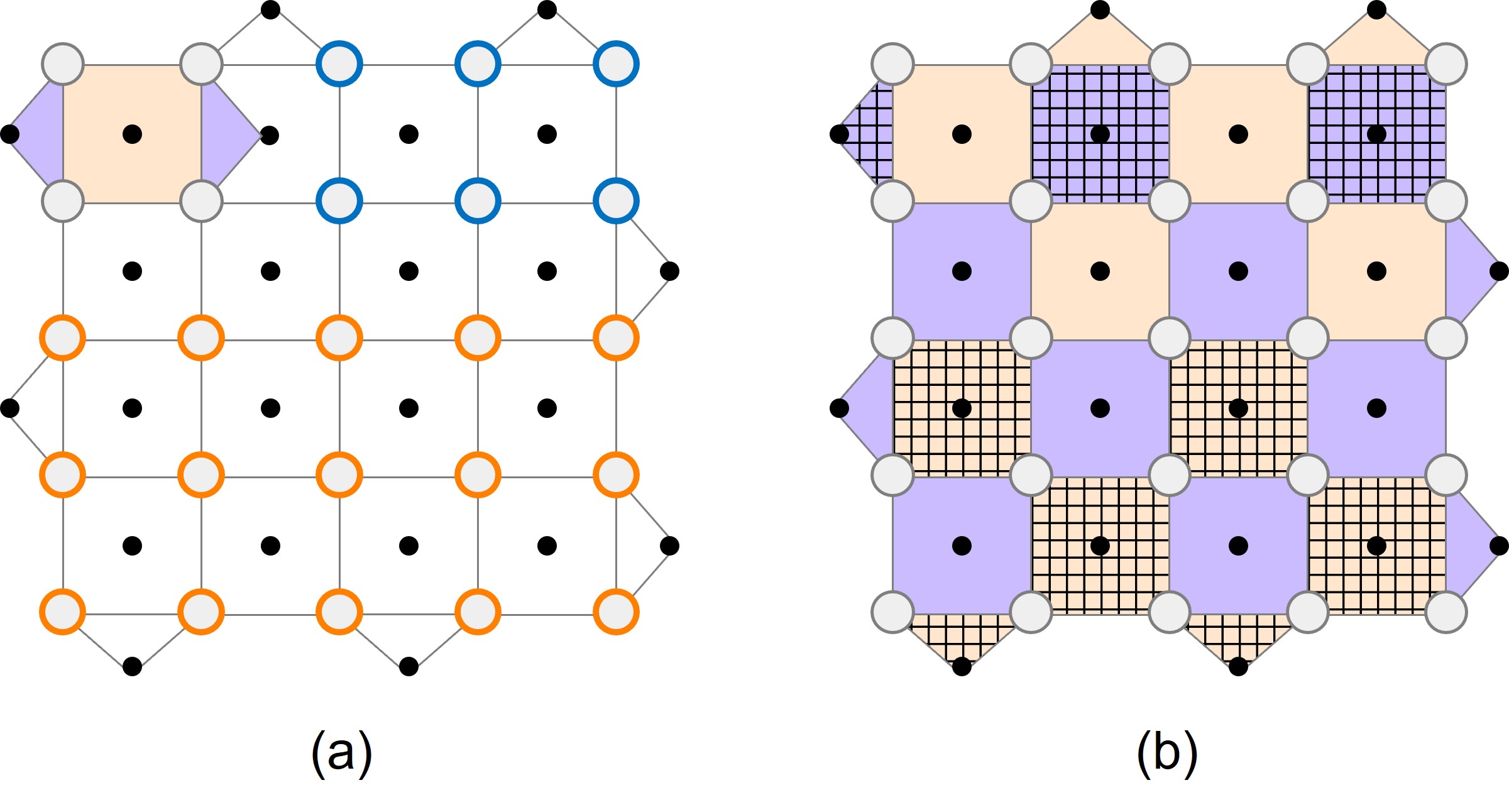}
  \caption{Expansion of the ancilla state to the $d = 5$ surface code patch. 
  (a) The ancilla state is prepared at the upper-left corner of the target patch. Other physical qubits are initialized to $\ket 0$ (blue circles) or $\ket +$ (orange circles). 
  (b) After the syndrome measurement, the hatched stabilizers have fixed eigenvalues determined by the initial configuration in an error-free case.  
   }
  \label{fig:expansion}
\end{figure}

Once we obtain the ancilla state $\ket{m_\theta}_L$ which passes the post-selection, 
we then expand it to the rotated surface code with an arbitrary code distance. 
We note that after the measurement circuit of Fig.~\ref{fig:422code_meascirc2}, the gauge DOF are fixed to the eigenstate of $G_Z = Z_0 Z_1$. 
In other words, it means that the post-selected state is stabilized by 
\begin{equation}
  S_X = X_0 X_1 X_2 X_3, S'_{Z} = Z_0 Z_1, S''_{Z} = Z_2 Z_3, 
\end{equation}
which is the smallest rotated planar surface code with $d = 2$. 
Therefore, its expansion to an arbitrary code distance patch can immediately achieved in a standard way in the lattice surgery\cite{Horsman_2012}. 
We show an example of the expansion to the $d = 5$ patch in Fig.~\ref{fig:expansion}.
The ancilla state $\ket{m_\theta}_L$ is prepared on a certain corner of the target patch, and other physical qubits in the target patch are initialized to $\ket 0$ (blue circles in Fig.~\ref{fig:expansion}) or $\ket +$ (orange circles in Fig.~\ref{fig:expansion}). 
Note that the success rate of the post-selection can be improved by performing the injection protocol in parallel using the empty space in the target patch. 
In this case, one picks up a successful ancilla state after the parallel injection and refreshes/initializes other physical qubits in the target patch.
Then, the syndrome measurement of the entire patch is performed twice. 
To obtain the ancilla state as clean as possible, we discard the expanded state if one of the following conditions is satisfied: 
(i) at least one of the syndromes is equal to $-1$, or
(ii) at least one of the bare syndromes whose value is determined by the initial configuration (hatched stabilizers in Fig.~\ref{fig:expansion} (2)) takes an unexpected value, although all syndromes are equal to $+1$. 
The state that passes the second post-selection is clean in detectable $O(p)$ error, and it can be consumed in the gate teleportation circuit of Fig.~\ref{fig:GT_circ} or the multi-Pauli rotation of Fig.~\ref{fig:mprot_circ}. 

Under the circuit-level noise model introduced later in Sec.~\ref{sect:result_surfacecode},
the logical error probability of the prepared ancilla state $\ket{m_\theta}_L$ behaves as follows:
\begin{eqnarray} \label{eq:ancilla_error_prob}
  P_{Z_L}(p) &=& 2p / 15 + O(p^2), \\
  P_{X_L}(p) &=& O(p^2),
\end{eqnarray}
whose details are discussed in Appendix~\ref{sect:appx1}.
Compared to the previous state injection protocols, our protocol achieves better precision even in more general situations. 
Let us briefly discuss the difference between other protocols and ours. 
In typical state injection protocols, 
one first prepares the ancilla state on a single physical qubit, and then it is injected into the encoded logical qubit. 
Since the first step directly suffers from the noisy qubit initialization and noisy single qubit operation, 
the injected state has a large logical error rate proportional to $p$. 
For example, the injection protocol proposed in Ref.~\cite{Li_2015} and its improved version of the rotated planar surface code~\cite{10.1145/3528416.3530237} show $P_L = 46p / 15 + O(p^2)$ and $P_L = 34p / 15 + O(p^2)$, respectively. 
In contrast, our protocol first generates an encoded qubit ${\ket +}_L$, 
then acts the logical $Z$ rotation gate $R_{Z_0 Z_2}(\theta)$ on the encoded qubit to generate the ancilla state. 
The $[[4,1,1,2]]$ subsystem code is $d = 2$ and most logical errors in our protocol occur at $O(p^2)$. 
Moreover, possible $O(p)$ logical errors are absorbed in part by the redundant gauge DOF and the circuit structure. 
Thus, our protocol achieves smaller error probability than the abovementioned protocols. 
Another recently proposed protocol is transversal injection~\cite{https://doi.org/10.48550/arxiv.2211.10046}, 
in which physical qubits are transversally initialized in a certain state before the encoding, and then a random state is injected depending on the initialization. 
By performing a post-selection, the authors report that the logical error rate of the injected state behaves as $P_L \approx 0.39 p$. 
Our protocol is advantageous for injecting a certain target state with a high accuracy because the injected state has no randomness and achieves better accuracy. 
Other approaches use distance-2 codes utilizing repetition code~\cite{pub.1147655643} and weight-2 hook propagation~\cite{https://doi.org/10.48550/arxiv.2302.12292}. 
However, the former method is weak for $O(p)$ bit-flip errors 
and the latter method still suffers from the propagation of a single qubit error that brings a large logical error rate when the single qubit error is not negligible. 
Because our protocol is robust against $O(p)$ bit-flip errors and does not assume that the single qubit errors are negligible, it is more versatile. 

Although our implementation has a small logical error rate, logical errors still occur at $O(p)$. 
Therefore, these remnant errors must be mitigated to obtain accurate results.
One possible mitigation technique applicable to the STAR architecture is the probabilistic noise cancellation (or the quasi-probability decomposition)~\cite{PhysRevLett.119.180509,PhysRevX.8.031027}. 
Let us consider the case where the noise channel is known as a simple phase-flip channel with an error probability $P$, 
\begin{equation} 
  {\mathcal E}(\rho) = (1-P) \rho + P Z \rho Z.
\end{equation}
In this case, we can explicitly construct an ``inverse error channel" ${\mathcal E}^{-1}$ as 
\begin{equation}
  {\mathcal E}^{-1}(\rho) = \frac{1-P}{1-2P} \rho - \frac{P}{1-2P} Z \rho Z, 
\end{equation} 
and can rewrite the noise-free (identity) channel as 
\begin{equation}
  {\mathcal I} = {\mathcal E}^{-1} {\mathcal E} = \gamma \left( (1-P) {\mathcal E} - P {\mathcal Z} {\mathcal E} \right), 
\end{equation}
where $\gamma = \frac{1}{1-2P}$ and ${\mathcal Z}$ is a Pauli $Z$ channel. 
Therefore, an error-free expectation value of a certain operator $M$ can be estimated by the noisy counterpart as 
\begin{equation} \label{eq:mitigation_qpm}
  \langle M \rangle_{\mathcal I} = \gamma \left( (1-P) \langle M \rangle_{\mathcal E} - P \langle M \rangle_{{\mathcal Z} {\mathcal E}} \right), 
\end{equation} 
where $\langle M \rangle_{\mathcal N} = {\rm tr} \left( M {\mathcal N}(\rho) \right)$. 
By performing Monte-Carlo sampling on an additional ${\mathcal Z}$ channel with a probability $P$, 
we can approximate Eq.~(\ref{eq:mitigation_qpm}) by averaging those samples with correct overall factors of $\pm \gamma$. 
The variance of the expectation value is amplified by a factor of $\gamma^2$ as seen in Eq.~(\ref{eq:mitigation_qpm}), so 
we need to generate $\gamma^2$ times more samples to suppress amplified statistical fluctuations. 

Returning to our analog rotation gate, 
its logical error channel is well described by the phase-flip channel with $P_L = 2p/15$ when $O(p^2)$ contributions are negligible,  
therefore the probabilistic error cancellation is applicable. 
This is an another benefit of our injection protocol. 
Note that the total step number of the RUS process varies in each sample, 
so that we cannot directly mitigate the errors of each RUS step. 
Instead, we consider the entire RUS process as a single noisy operation and mitigate its error. 
The logical $Z$ error probability of the entire RUS process is given as 
\begin{equation} \label{eq:RUS_zerrorrate}
  P = \sum_{n=1}^{\infty} \frac{1}{2^n} P_{Z, n}, 
\end{equation}
where $P_{Z, n}$ is an error probability when the RUS process is completed by $n$-th step: 
\begin{eqnarray}
  P_{Z, n} &=& \sum_{m=1}^{ \lfloor \frac{n+1}{2} \rfloor } \begin{pmatrix} n \\ 2m-1 \end{pmatrix} P_{Z, 1}^{2m-1} (1-P_{Z, 1})^{n-2m+1} \nonumber \\ 
  &=& n P_{Z, 1} + O(P_{Z,1}^2), \\
  P_{Z, 1} &=& 2p / 15. 
\end{eqnarray}
Since $O(P_{Z,1}^2) = O(p^2)$ can be neglected, Eq.(\ref{eq:RUS_zerrorrate}) becomes
\begin{equation}
  P = \sum_{n=1}^{\infty} \frac{1}{2^n} P_{Z, n} = P_{Z,1} \sum_{n=1}^{\infty} \frac{n}{2^n} = 2 P_{Z,1}. 
\end{equation}
Therefore, we can mitigate the phase-flip error of the entire RUS process by the probabilistic error cancellation with $P=2P_{Z,1}$ 
by an additional sampling overhead of $\gamma^2 = (\frac{1}{1-2P})^2 \approx e^{8P_{Z,1}}$ ($P \ll 1$). 
When we perform $N$ analog rotations in a circuit and want to mitigate their noise, 
we can immediately extend the discussion by assuming that $N$ noise channels are independent. 
In such a case, the sampling overhead is modified to $\gamma^{2N} \approx e^{8P_{Z,1}N}$. 

\begin{figure}[tbp]
  \centering
  \includegraphics[width=80mm, clip]{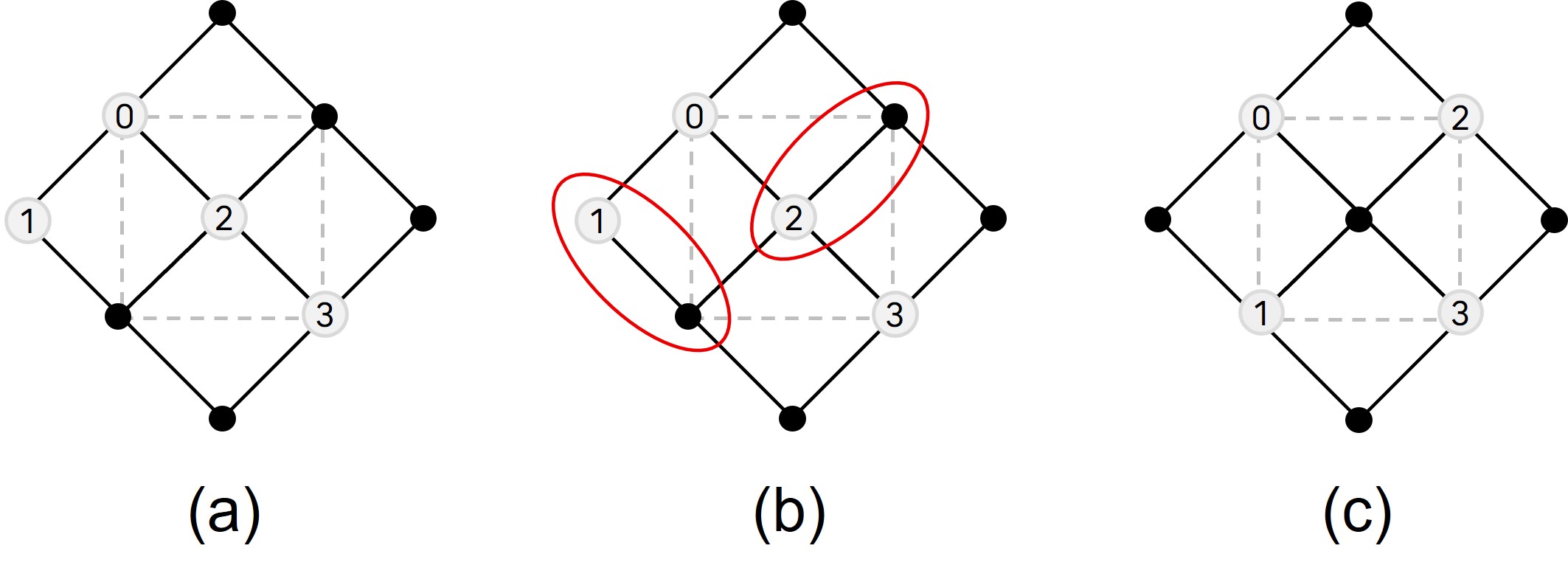}
  \caption{Example of the insertion of SWAP gates. 
  (a) Initial physical qubit arrangement. 
  The circuit in Fig.~\ref{fig:422prep_circ} is performed in this arrangement. 
  (b) Physical qubits 1 and 2 are then swapped to the neighboring measurement qubits (red circled pairs). 
  (c) Final arrangement after SWAP operations. 
  }
  \label{fig:NN_SWAP}
\end{figure}

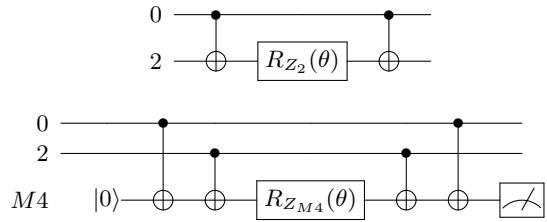
\begin{figure}[tbp]
  \centering
  \mbox
  {
  \Qcircuit @C=1.3em @R=1.0em {
    \lstick{0} & \ctrl{1} & \qw                & \ctrl{1} & \qw \\
    \lstick{2} & \targ    & \gate{R_{Z_2} (\theta)} & \targ    & \qw 
  }
  } 
  \\ \vspace{5mm}
  \mbox
  {
  \Qcircuit @C=1.3em @R=1.0em {
    \lstick{0} & \qw            & \ctrl{2} & \qw      & \qw                        & \qw      & \ctrl{2} & \qw \\
    \lstick{2} & \qw            & \qw      & \ctrl{1} & \qw                        & \ctrl{1} & \qw      & \qw \\
    \lstick{M4}& \push{\ket{0}} & \targ    & \targ    & \gate{R_{Z_{M4}} (\theta)} & \targ    & \targ    & \meter
  }
  }
  \caption{Examples of a circuit for indirectly performing $R_{Z_0 Z_2}(\theta)$ operation.
  (Upper) $R_{Z_0 Z_2}(\theta)$ can be performed using a single qubit rotation $R_{Z_2}(\theta)$ and CNOTs. The logical error rate degrades to $P_L = 9p / 15 + O(p^2)$ in this circuit since there are more error patterns generating the logical $Z$ error. 
  (Lower) An alternative $R_{Z_0 Z_2}(\theta)$ operation circuit utilizing another ancilla qubit $M4$. The final $Z$ measurement of the ancilla $M4$ is necessary to detect $O(p)$ $X$ errors which flip $\theta$ to $-\theta$. The logical error rate becomes $P_L = 7p / 15 + O(p^2)$ in this circuit.}
  \label{fig:zzrot_alt}
\end{figure}

Finally, we briefly discuss how the restrictions of real devices affect on our protocol. 
In real quantum devices, there are several restrictions on the connectivity between physical qubits and native gate sets. 
Regarding the connectivity restriction, for example, 
the superconducting qubits can only interact with their nearest neighbors. 
If we consider this restriction in a qubit arrangement such as that of Fig.~\ref{fig:422code_meascirc1}, 
only qubits connected by black solid lines interact with each other; 
thus, the CNOT operation and $R_{Z_0Z_2}(\theta)$ in the circuit of Fig.~\ref{fig:422prep_circ} cannot be directly performed. 
In this situation, one can resolve the connectivity problem by additionally inserting SWAP gates in the circuit. 
We show a simple example to perform our state injection circuit with SWAP gates in Fig.~\ref{fig:NN_SWAP}.
Inserted SWAP gates also introduce additional two-qubit errors,
but they occur on certain pairs of measurement and physical qubits, so they do not lead to logical errors at $O(p)$. 
Therefore, the  performance of our protocol at $O(p)$ can be maintained. 
Regarding the restriction of native gates, our assumption that the $R_{Z_0Z_2}(\theta)$ gate can be directly applied may become invalid in some cases. 
For typical ion trap devices and superconducting devices, this assumption is valid since the $X \otimes X$ rotation and the $Z \otimes X$ rotation (the cross resonance gate) can be directly implemented respectively. 
If such gates are not supported, the indirect implementations of the $R_{Z_0Z_2}(\theta)$ gate shown should be used. 
One straightforward example is the circuit shown in Fig.~\ref{fig:zzrot_alt} (Upper). 
In this example, the logical error rate of the injected state degrades to $P_L = 9p / 15 + O(p^2)$ under the circuit-level noise model 
since the single $Z$ errors and their propagation through the second CNOT gate lead to additional undetectable logical $Z$ errors. 
If an ancilla qubit is available for this operation, we can employ another circuit where the logical error rate behaves as $P_L = 7p / 15 + O(p^2)$ [Fig.~\ref{fig:zzrot_alt} (Lower)]. 
The ancilla qubit is measured by $Z$ basis after the operation to detect $X$ errors which flip the rotation angle $\theta$ to $-\theta$. 
If any $X$ error is detected, the injection protocol will be restarted. 
In the following resource estimation discussed in Sect.~\ref{sect:resource_estimation}, we directly perform the circuit in Fig.~\ref{fig:422prep_circ} 
without specifying any device restriction.

\section{Logical qubit arrangement}
\label{sec:lqa}
As already mentioned in Sec.~\ref{sect:latticesurgery}, mainly two schemes are used to perform quantum computations by the lattice surgery: 
(I) Original quantum circuits are computed by using explicit logical Clifford gates and single qubit rotation gates $R_Z(\theta)$, and 
(II) Quantum circuits are converted to alternative forms comprising multi-Pauli rotation gates and multi-Pauli measurements beforehand, then the converted circuits are computed.
The former case is suitable for calculating quantum circuits which have a high parallelism of the rotation gates since the rotation gate only acts on a single logical qubit and can be performed parallelly. 
A major drawback is that we must implement costly logical Clifford operations, which need $2d$ or $3d$ rounds of the syndrome measurement and additional ancilla patch. 
In the latter case, on the other hand, the number of physical qubits required is small because it does not need an ancilla region for the explicit Clifford gates. 
Instead, multi-Pauli rotation gates are difficult to parallelize because of the anti-commutation relations between them. 
Therefore, we can say that the latter scheme is suitable for calculating quantum circuits that contain the rotation gates sparsely or have a small parallelism of the rotation gates. 

The arrangement of the logical qubit patches strongly depends on these schemes. 
Moreover, a trade-off relationship holds between the efficiency of the number of the logical qubits and the execution time of the operations in general. 
To minimize unnecessary overheads, one needs to find an optimal arrangement for an input quantum circuit. 
Additionally, the input circuit possibly must be converted into a suitable form before the determination of the optimal arrangement. 
A logical qubit arrangement optimizer and circuit compiler are mandatory for maximizing the computational power of the STAR architecture, but it is beyond the scope of this paper. 
Here, we only illustrate some typical arrangements of the logical qubit in both schemes. 
The development of a circuit compiler and arrangement optimizer for the STAR architecture 
is one of the most important future studies. 

\begin{figure}[tbp]
  \centering
  \includegraphics[width=30mm, clip]{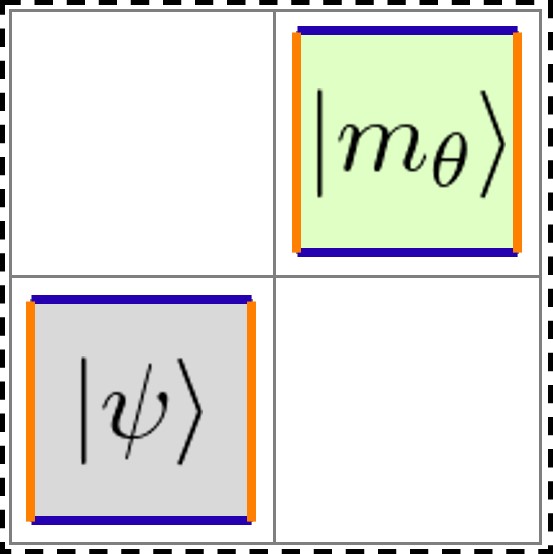}
  \caption{Data unit that carries a data logical qubit and ancilla state for rotation gates. 
  Actual data are carried by the logical patch labeled as $\ket \psi$, 
  and the other patch labeled as $\ket{m_{\theta}}$ is the ancilla state for the rotation gate.}
  \label{fig:scheme1_unit}
\end{figure}

\begin{figure}[tbp]
  \centering
  \includegraphics[width=80mm, clip]{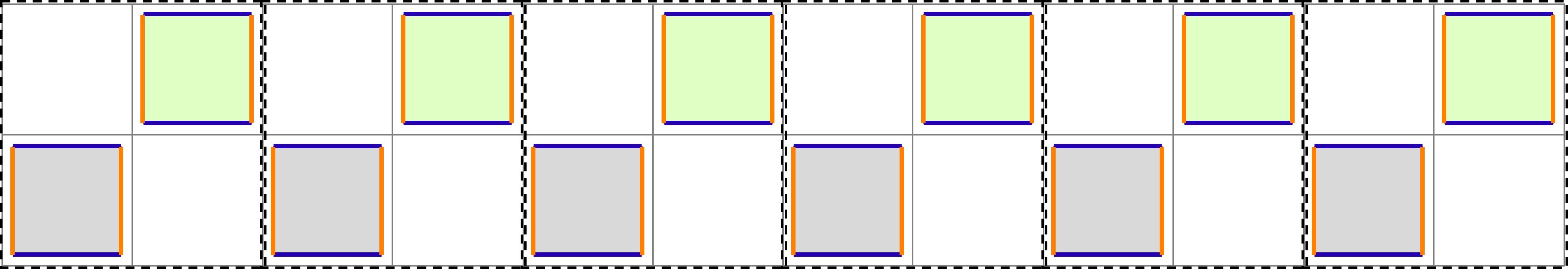}
  \caption{Example of the qubit arrangement in scheme (I) with $n = 6$. 
  Data units are arranged in a row. }
  \label{fig:scheme1_arrangement}  
\end{figure}

\begin{figure}[tbp]
  \centering
  \includegraphics[width=80mm, clip]{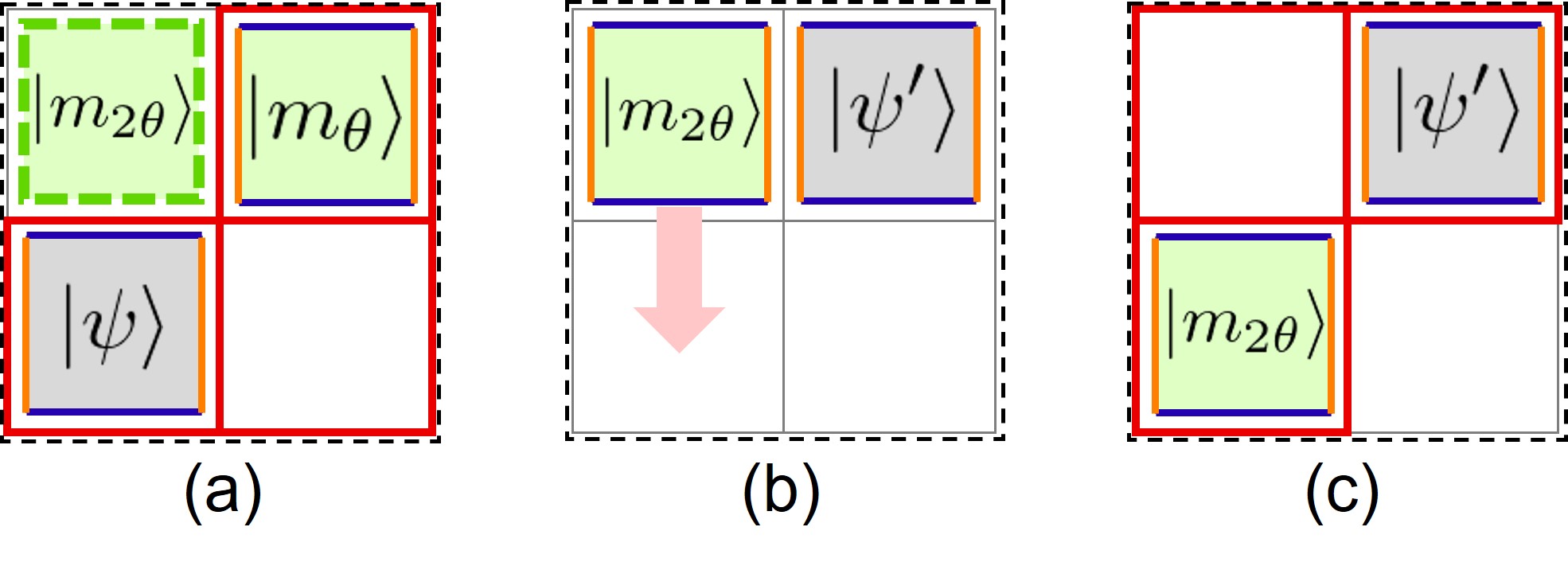}
  \caption{RUS protocol within the data unit. 
  (a) Gate teleportation circuit of Fig.~\ref{fig:GT_circ} is computed using three patches, where $\ket{m_{\theta}}$ and $\ket{\psi}$ are the control qubit and target qubit of the CNOT gate, respectively (red square). 
  During the computation, we can prepare the ancilla state $\ket{m_{2\theta}}$ for the next RUS step in the lower-right patch (green dashed square). 
  If the output state $\ket {\psi'}$ is not correct, (b) the prepared ancilla state moves to the next patch,  
  then (c) the next step of the RUS protocol begins. 
  We can also prepare $\ket{m_{4\theta}}$ using the upper-left patch. }
  \label{fig:scheme1_RUS}
\end{figure}

\begin{figure}[tbp]
  \centering
  \includegraphics[width=80mm, clip]{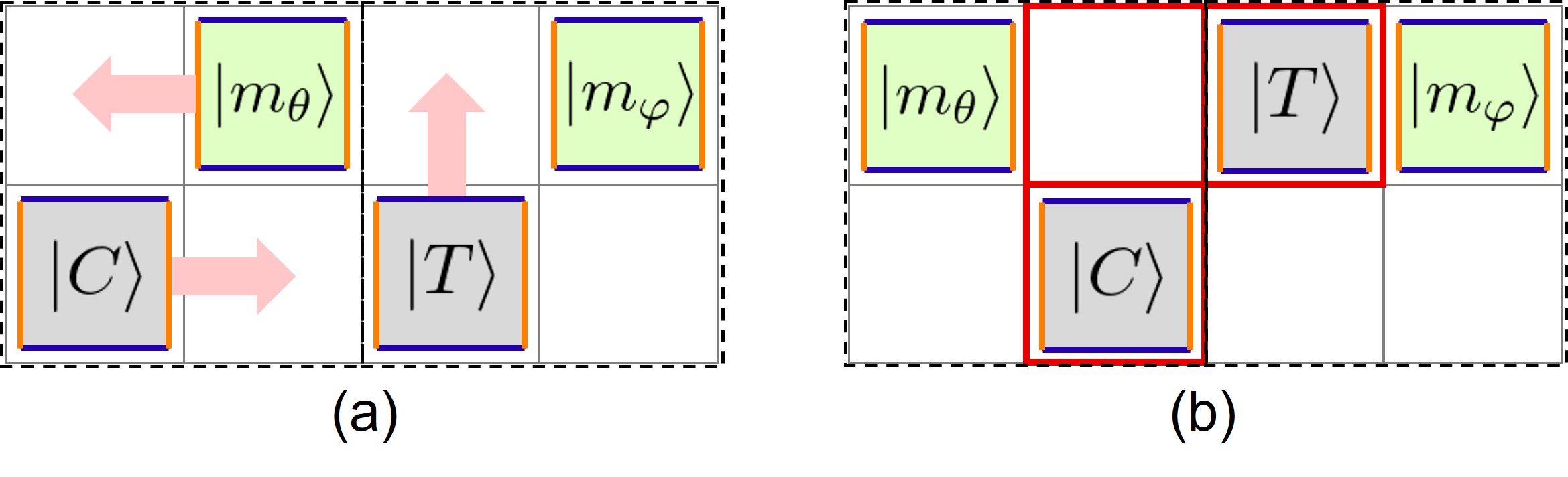}
  \caption{Direct logical CNOT operation between neighboring data units. 
  (a) The logical patches in these units first move to the appropriate positions, and then (b) the CNOT operation is performed (red square). 
  }
  \label{fig:scheme1_nncnot}
\end{figure}

\begin{figure}[tbp]
  \centering
  \includegraphics[width=80mm, clip]{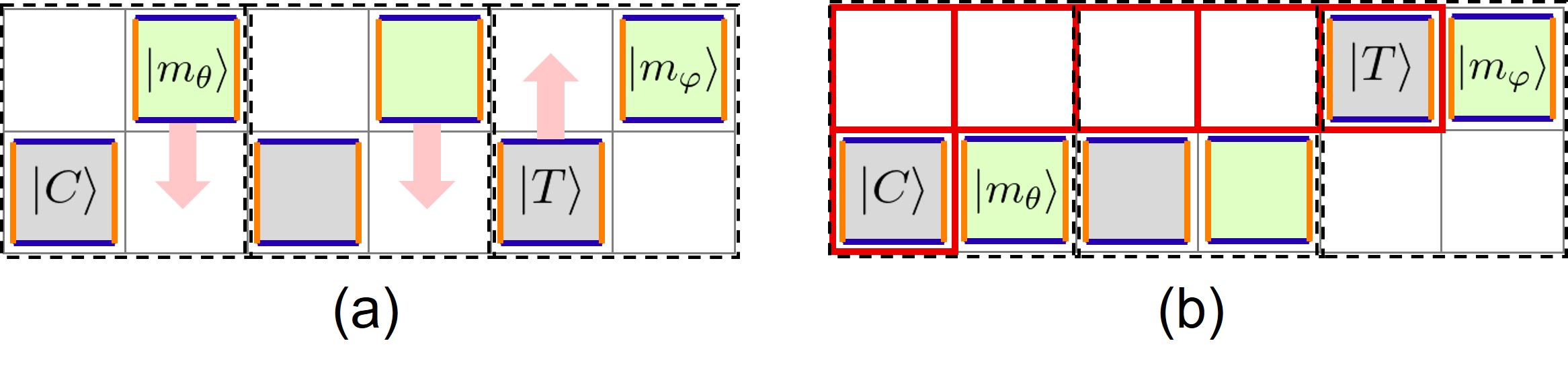}
  \caption{Remote logical CNOT operation. 
  (a) Logical patches move to the appropriate positions. 
  (b) Then the CNOT operation is performed using a long ancilla patch (red square). }
  \label{fig:scheme1_remotecnot}
\end{figure}

In scheme (I), the ancilla state $\ket{m_\theta}$ should be prepared in parallel for each data logical qubit to maximize the merit of the parallelism of rotation gates. 
Given that the gate teleportation circuit of Fig.~\ref{fig:GT_circ} needs at least three logical patches because of the logical CNOT operation, 
it is better to group four logical patches as a unit, which carry a data logical qubit and an ancilla qubit for parallel rotation gates, 
as shown in Fig.~\ref{fig:scheme1_unit} (note that the same structure has been proposed in Ref.~\cite{Lao_2019}). 
Figure~\ref{fig:scheme1_arrangement} exemplifies the arrangement based on this unit. 
This example requires at least $4n$ logical patches to allocate $n$ data logical qubits. 
We can perform the RUS protocol within the unit as shown in Fig.~\ref{fig:scheme1_RUS}. 
Because a single patch is free during the gate teleportation circuit, 
we can prepare the ancilla state needed for the next RUS step with a small overhead (dashed green square in Fig.~\ref{fig:scheme1_RUS} (a)). 
Additionally, the patch rotation after the logical $H$ operation (Fig.~\ref{fig:surgery_h}) can be done using two patches in the unit.  
The logical CNOT operation can be directly applied to neighboring units with additional patch movements. 
Moreover, remote CNOT operations between distant units can be realized using the ancilla region. 
We show examples of those logical CNOT operations in Figs.~\ref{fig:scheme1_nncnot} and~\ref{fig:scheme1_remotecnot}.  
Note that one cannot perform some remote CNOT operations parallelly in the architecture of Fig.~\ref{fig:scheme1_arrangement} 
because their ancillae for them cannot overlap. 
To minimize such conflicts, it is better to optimize the mapping of the quantum circuit.  

\begin{figure}
    \centering
    \includegraphics[width=30mm, clip]{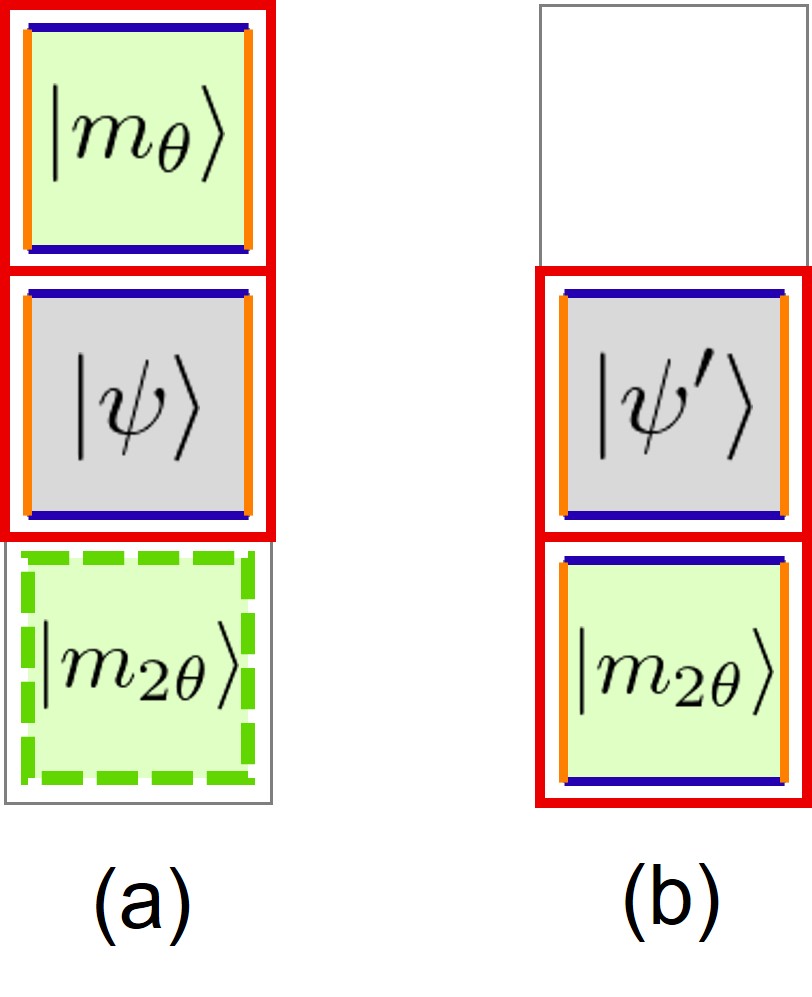}
    \caption{RUS protocol based on the multi-Pauli rotation circuit of Fig.~\ref{fig:mprot_circ}. 
    (a) A single $Z$ rotation circuit is implemented by a $Z \otimes Z$ measurement between the target and ancilla states (red square). 
    During the operation, the other logical patch can be used to prepare the next ancilla state (green dashed square). 
    (b) After the first RUS step, the output state $\ket{\psi'}$ remains; thus, the next RUS step can start immediately. }
    \label{fig:scheme1_RUS_alt}
\end{figure}

\begin{figure}
    \centering
    \includegraphics[width=60mm, clip]{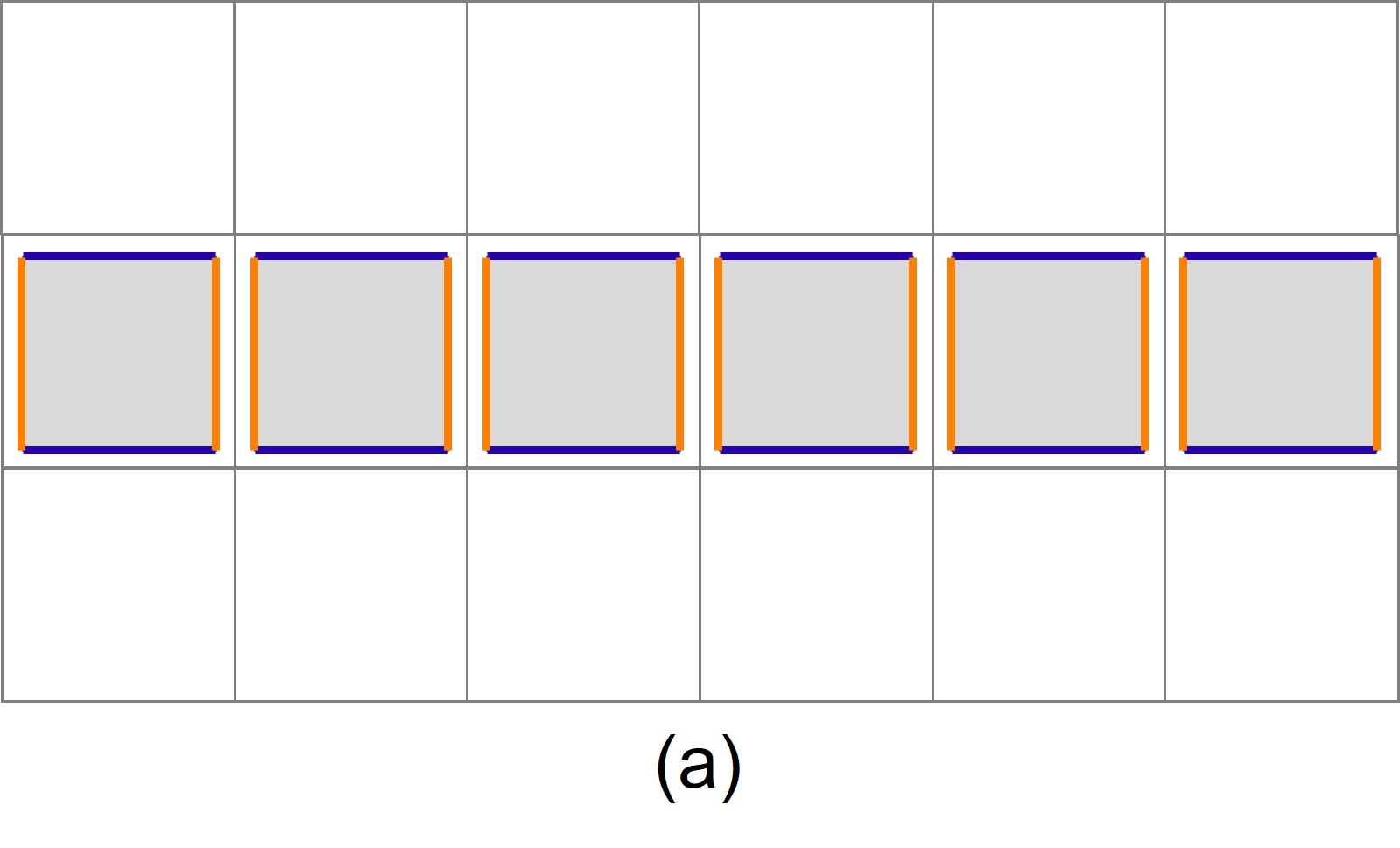}
    \includegraphics[width=60mm, clip]{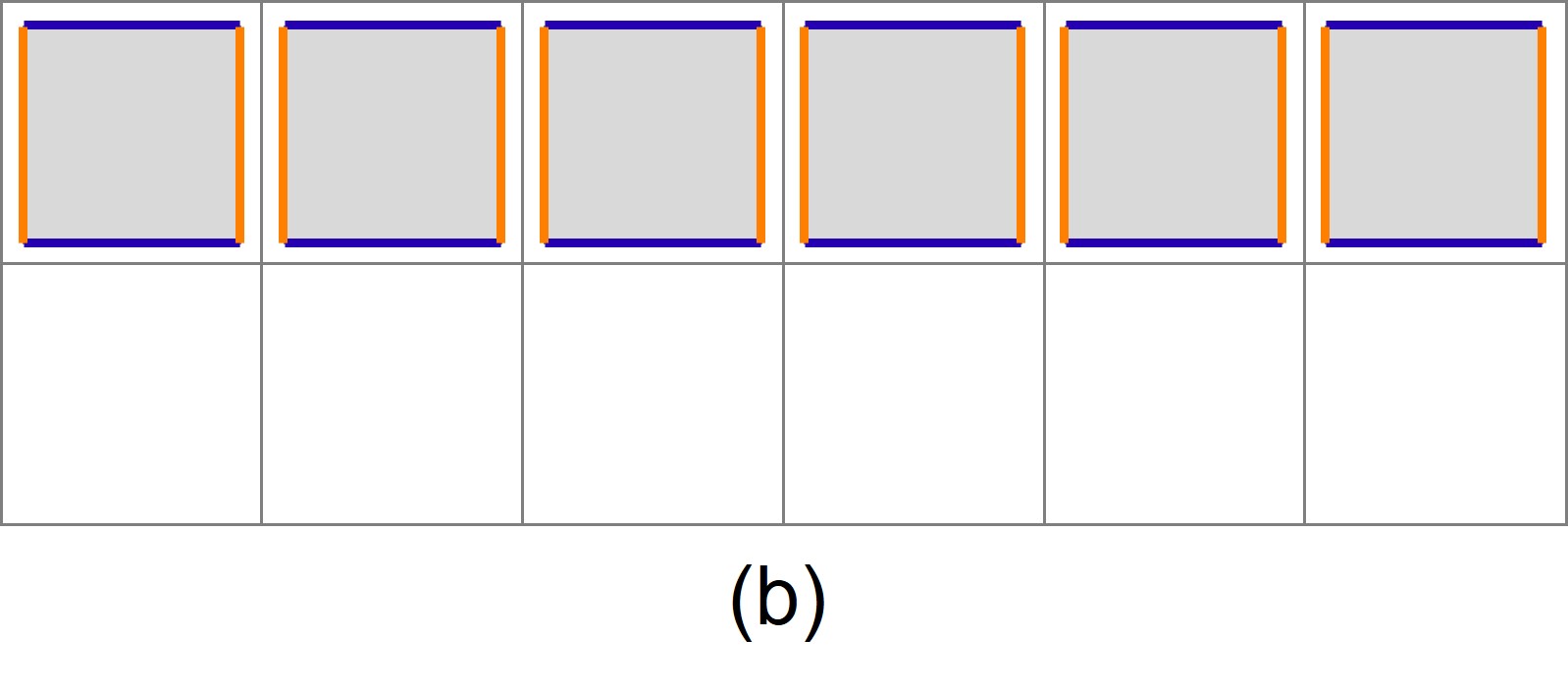}
    \caption{Typical qubit arrangement in scheme (I) based on the circuit of Fig.~\ref{fig:mprot_circ} with $n = 6$. 
    (a) Arrangement requiring $3n$ patches. 
    Each column of three patches constructs a unit to implement the RUS protocol, and they are arranged in a row. 
    An ancilla region is used not only to parallel $Z$ rotations but also to perform other operations, such as logical CNOT operations and patch deformations. 
    (b) Minimum arrangement requiring $2n$ patches. 
    Although this arrangement cannot hide the overhead of the state injection during the RUS protocol, it only require $2n$ patches to allocate $n$ data logical qubits.}
    \label{fig:scheme1_arrange_alt}
\end{figure}

The requirement of the $4n$ logical patch discussed above may be somewhat large. 
Fortunately, 
we can use the multi-Pauli measurement-based rotation circuit of Fig.~\ref{fig:mprot_circ} 
instead of the gate teleportation circuit of Fig.~\ref{fig:GT_circ} to reduce the number of logical qubit patches. 
If we use the circuit of Fig.~\ref{fig:mprot_circ} for a single qubit rotation, 
we must measure the $Z \otimes Z$ operator over the target logical state and the ancilla state, 
but it can be immediately performed by the $Z$-boundary merging and splitting of these states~\cite{Litinski2019gameofsurfacecodes}. 
Furthermore, unlike the circuit of Fig.~\ref{fig:GT_circ}, the target logical state does not move after the rotation; 
thus, no unnecessary overhead is needed to bring it back to the correct place. 
Therefore, in this case, we can perform the RUS protocol by a unit of three logical patches contacting each other on $Z$-boundaries as shown in Fig.~\ref{fig:scheme1_RUS_alt}. 
During a single rotation performed by two of the three patches, 
the other patch can prepare the ancilla state for the next rotation. 
Because the target logical patch remains at the same position after the rotation, the next rotation step can immediately start. 
Figure~\ref{fig:scheme1_arrange_alt} (a) provides a typical arrangement in this case, which requires $3n$ logical patches to allocate $n$ data logical qubits. 
One can perform CNOT operations between logical patches and patch deformations in the same way as in the previous arrangement of Fig.~\ref{fig:scheme1_arrangement} using the ancilla region.  
Note that if the overhead of the state injection does not need to be hidden, 
then the number of the logical patches can be further reduced to $2n$, as shown in Fig.~\ref{fig:scheme1_arrange_alt} (b). 
Although we mainly consider the case in which all logical data patches have identical unit structures, 
they can be mixed to minimize the computational overhead. 

\begin{figure}[bp]
  \centering
  \includegraphics[width=50mm, clip]{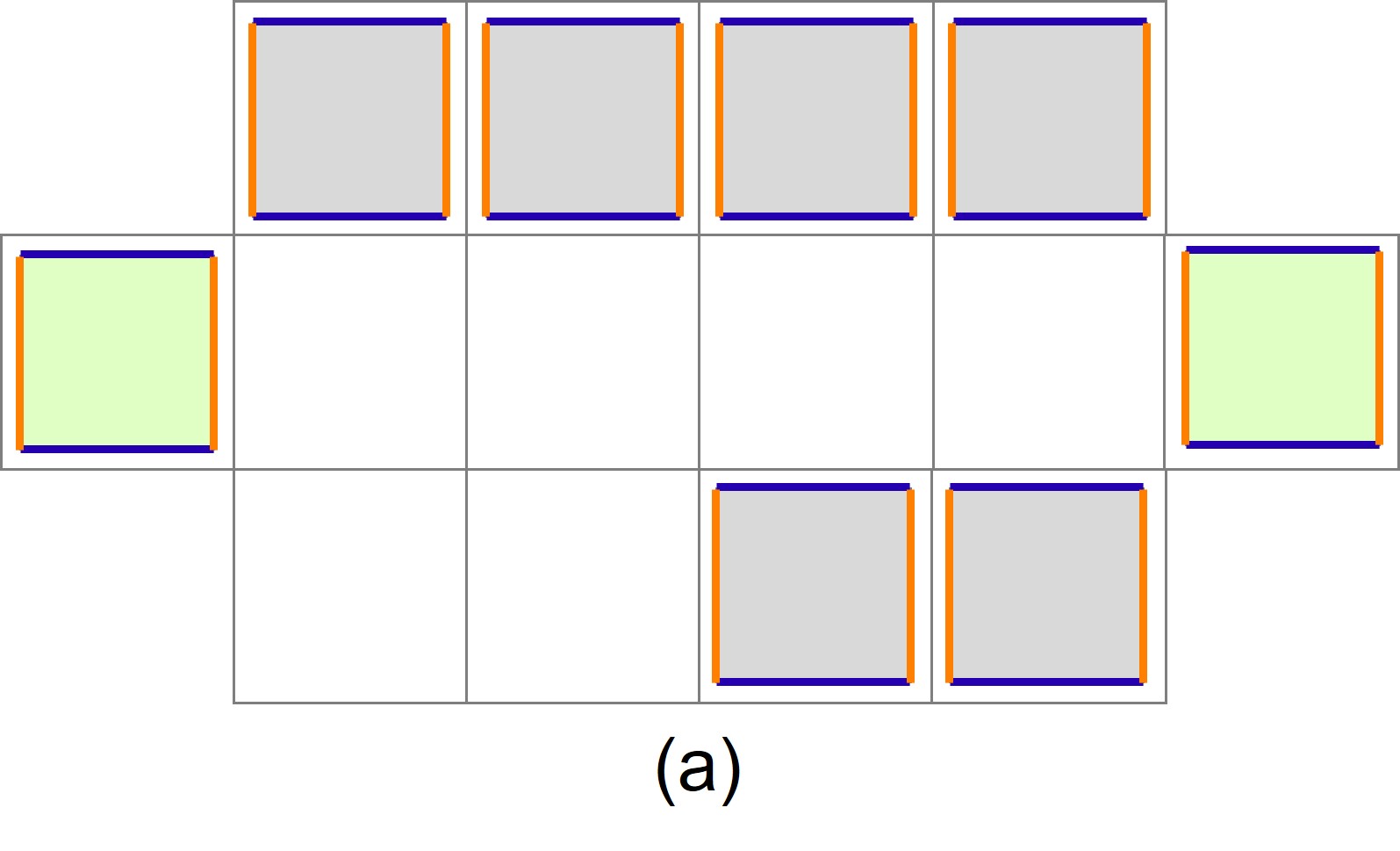} \\ \vspace{5mm}
  \includegraphics[width=80mm, clip]{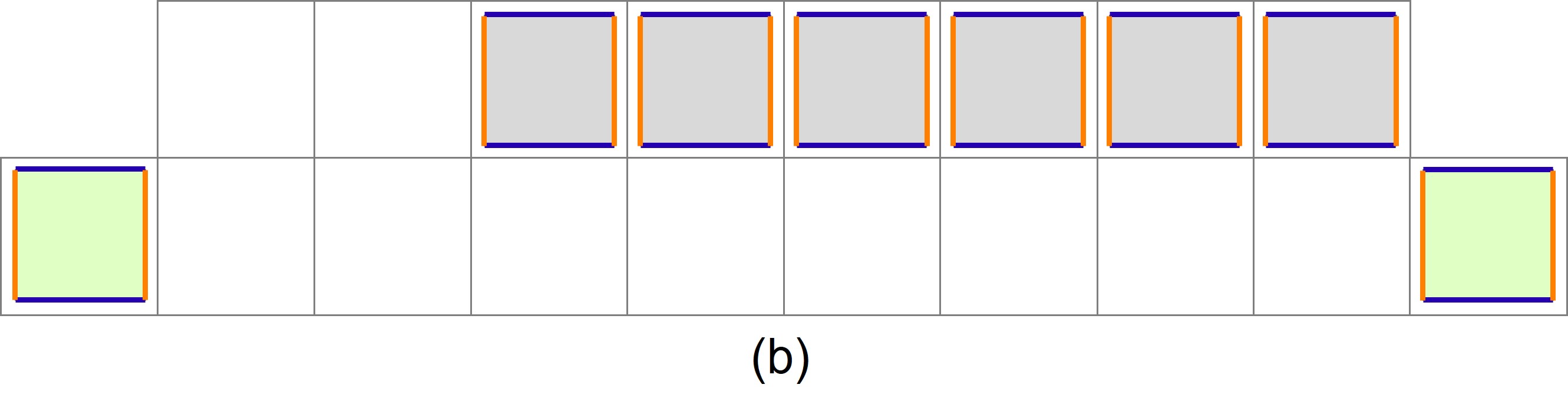}
  \caption{Example of the qubit arrangement in scheme (II), based on the data blocks discussed in Ref.~\cite{Litinski2019gameofsurfacecodes}. 
  (a) Compact block for $n = 6$. 
  Gray patches represent data logical qubits. 
  We allocate two patches to prepare the ancilla state (green) to reduce its additional overhead. 
  (b) Intermediate block for $n = 6$.}
  \label{fig:scheme2_arrangement}
\end{figure}

The prototypical logical qubit arrangement in scheme (II) was proposed in detail in Ref.~\cite{Litinski2019gameofsurfacecodes}, and we only briefly introduce it here. 
Since the early-FTQC device has a limited number of physical qubits, 
the compact and intermediate block~\cite{Litinski2019gameofsurfacecodes} are suitable for our purpose. 
Figure~\ref{fig:scheme2_arrangement} shows typical examples of each cases. 
In these example, we assume that the multi-Pauli rotation gates are sequentially performed. 
We allocate two additional patches for the injection of the ancilla state $\ket{m_{\theta}}$ to hide the overhead of the state injection behind the execution time of a single RUS step by consuming and generating the ancilla states consecutively. 
The minimum construction using the compact and intermediate blocks requires $1.5n + 5$ and $2n + 6$ logical patches to allocate $n$ data logical qubits, respectively. 

\section{Performance of the STAR architecture} \label{sect:resource_estimation}
To estimate the performance of our proposal quantitatively, 
we perform numerical simulations on the error correction of the surface code patch (related to the orange square part in Fig.~\ref{fig:overview}) and the ancilla state injection (related to the blue square part in Fig.~\ref{fig:overview}).
In this section, 
we show the results of these simulations and 
estimate the computational resources available for early-FTQC devices based on these results. 
We also briefly discuss possible applications of the STAR architecture based on the estimation. 

\subsection{Logical error probability of the rotated surface code patch} \label{sect:result_surfacecode}
In the simulation of the error correction of the rotated surface code, 
we assume that Hadamard and CNOT, initialization to $\ket 0$, and measurement in the $Z$ basis are available as physical qubit operations, 
so that we employ the depth 8 measurement circuit of Fig.~\ref{fig:meas_circ}. 
Noise processes are simulated by the circuit-level noise model, 
in which all operations on physical qubits suffer from errors: 
Noisy qubit initialization and measurement flip to an orthogonal state with a probability $p$, and 
noisy Hadamard and CNOT gates are simulated by ideal gate operations followed by the depolarizing noise channels, 
\begin{equation} \label{eq:single_depolarizing}
  {\mathcal E}_{\rm single}(\rho) = (1-p) \rho + \frac{p}{3} \left( X \rho X + Y \rho Y + Z \rho Z \right), 
\end{equation}
and
\begin{eqnarray} \label{eq:double_depolarizing}
  && {\mathcal E}_{\rm double}(\rho) = \left(1-\frac{16}{15}p \right) \rho 
    + \frac{p}{15}
\sum _{E \in \{ I,X,Y,Z \}^{\otimes 2} } E \rho E , \nonumber \\
\end{eqnarray}
respectively. 
Noisy identity gates are inserted whenever physical qubits are idle. 
We assume that all errors occur with a common probability $p$.
All measurement circuits are performed in parallel and repeated $d$ times to treat measurement errors. The last measurement round is performed ideally.
For the decoding, we employ PyMatching~\cite{pymatchingv2}, an open-source Python/C++ library, 
to implement the MWPM algorithm. 
We consider hook error edges in the construction of the decoder graph 
to decode errors correctly up to $O(p^{\lfloor \frac{d-1}{2}\rfloor})$. 

\begin{figure*}[tbp]
  \centering 
  \includegraphics[width=135mm, clip]{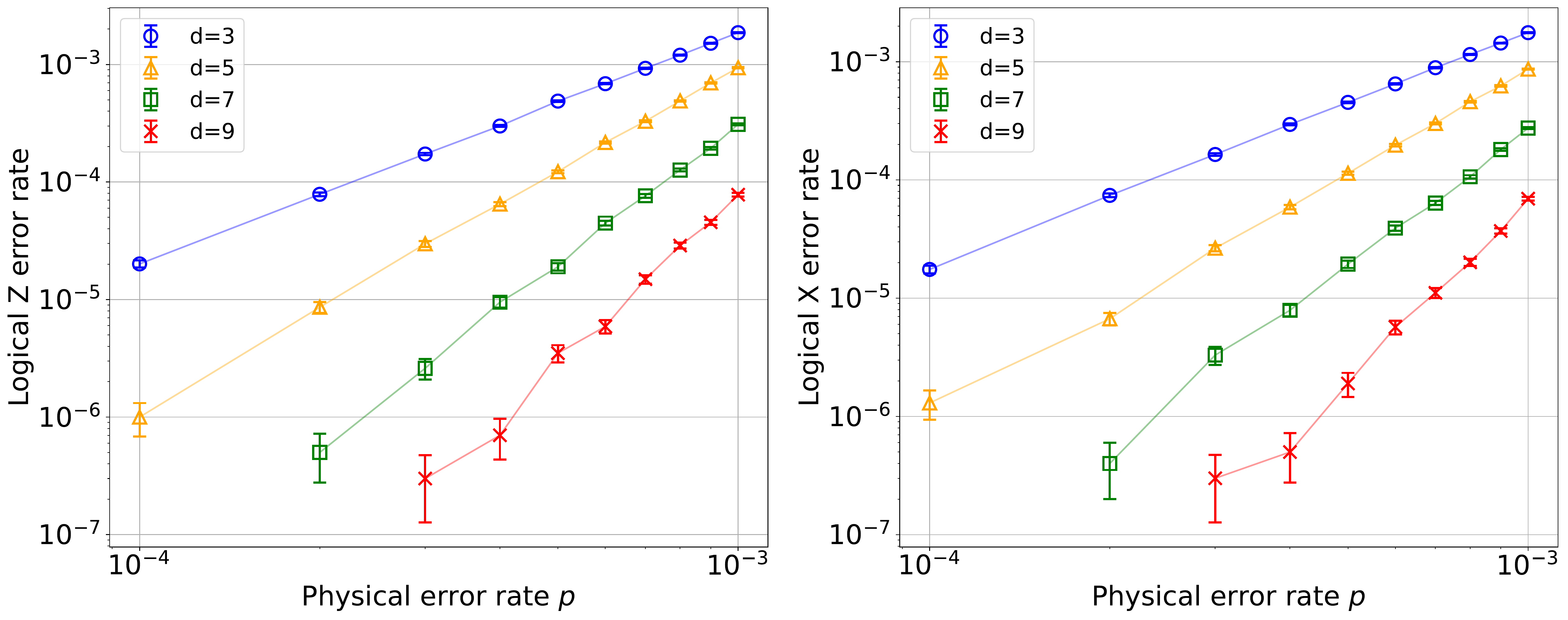}
  \caption{Logical $Z$ (left) and $X$ (right) error rates of the rotated surface code patch. 
  The error bars indicate $\pm 1 \sigma$ statistical errors estimated by a standard deviation of the Monte Carlo samples.}
  \label{fig:surfacecode_logerror}  
\end{figure*}

\begin{figure*}[tbp]
  \centering
  \includegraphics[width=135mm, clip]{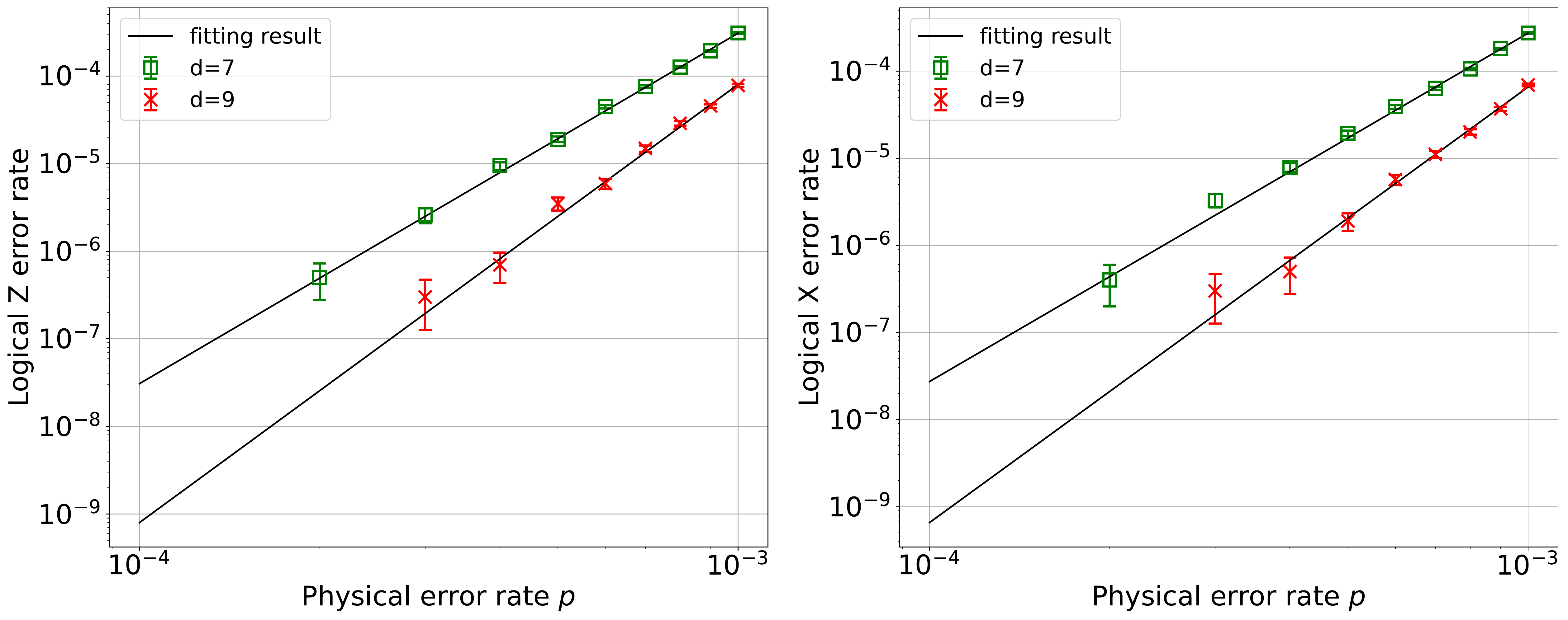}
  \caption{Fitting results of the logical error rates with $d = 7, 9$. 
  The scaling of Eq.~(\ref{eq:scaling_func}) with the mean values of the optimized parameters are shown as solid black lines.}
  \label{fig:surfacecode_logerror_fit}  
\end{figure*}

Logical error rates $P_{L, i} (i = Z, X)$ are determined by $10^7$ Monte Carlo samples for each physical error rate $p$. 
For a resource estimation under a limited number of physical qubits available in early-FTQC era, 
we consider small code distances of up to $d = 9$ and a physical error rate of $p \in [10^{-4}, 10^{-3}]$.
Figure~\ref{fig:surfacecode_logerror} shows the resultant logical error rates obtained in our simulation. 
Because obtained data seems to behave linearly in the log-log plot as seen in Fig.~\ref{fig:surfacecode_logerror}, 
we can expect that the $p$ dependence of the logical error rate is well described by 
\begin{equation} \label{eq:scaling_func}
  P_{L, i}(p) = C_{i} \left( \frac{p}{p_{th, i}} \right)^{\frac{d+1}{2}} \qquad (i = Z, X),
\end{equation} 
where $C_{i}$ and $p_{th, i} (i = Z, X)$ are constant parameters, within the range of the physical error rate we consider. 
We determine those parameters by fittings using the numerical results of $d = 7, 9$. 
We show the optimized parameters and the behaviors of Eq.~(\ref{eq:scaling_func}) with the optimized parameters 
in Tab.~\ref{tab:global_fit} and Fig.~\ref{fig:surfacecode_logerror_fit}, respectively.
As expected, the numerical data are well fitted by the function of Eq.~(\ref{eq:scaling_func}). 
We also observe that the obtained threshold value $p_{th, X}$ is larger than $p_{th, Z}$, 
which is a well-known behavior resulting from the circuit asymmetry of Fig.~\ref{fig:meas_circ}~\cite{PhysRevA.89.022321}. 
In the later resource estimation, we employ Eq.~(\ref{eq:scaling_func}) with the mean values of the optimized parameters (black solid lines in Fig.~\ref{fig:surfacecode_logerror_fit}).
\begin{table}[tbp]
  \centering
  \caption{Optimized parameters of Eq.~(\ref{eq:scaling_func}) by the fitting. 
  Statistical errors are estimated by the jackknife method with a bin size of $10^4$.}
  \label{tab:global_fit}
  \begin{tabular}{cccc}
    \hline \hline
    $C_Z$ & $p_{th, Z}$ & $C_X$ & $p_{th, X}$ \\ \hline
    0.0679(76) & 0.00385(10) & 0.0819(97) & 0.00416(12) \\
    \hline \hline
  \end{tabular}  
\end{table}

\subsection{Logical error probability of the ancilla state} \label{sect:result_ancprep}
In this study, we simulate the entire process of the state injection protocol discussed in Sec.~\ref{sect:state_prep}. 
Because the target patch after the expansion contains many physical qubits, 
the simulation is performed based on the stabilizer formalism~\cite{PhysRevA.70.052328}. 
The stabilizer simulation does not support non-Clifford gates; thus, we take $\theta = 0$. 
We can justify this assumption as follows: 
First, in this setup, we cannot estimate the logical $X$ error rate because the prepared state is now $R_{Z}(0)\ket + = \ket +$, on which the logical $X$ operator does nothing. 
As discussed in Appendix~\ref{sect:appx1}, however, the logical $X$ error only occurs at $O(p^2)$ and can be neglected if we are interested in the leading $O(p)$ logical error rate. 
Second, the elimination of $R_{Z_0 Z_2}(\theta)$ ignores some error propagation processes occurring at $O(p)$, such as $R_{Z_0 Z_2}(\theta) X_0 = X_0 R_{Z_0 Z_2}(-2\theta) \cdot R_{Z_0 Z_2}(\theta)$, but those error always occurs together with a single Pauli $X$ operator, which is detectable in the following post-selection. 
Therefore, the elimination does not modify the leading $O(p)$ logical error rate. 
We employ the circuit-level noise model and the same assumption on the fundamental operations 
as the simulation of the surface code patch already discussed in Sec.~\ref{sect:result_surfacecode}.
Syndrome measurements are performed using the circuits of Figs.~\ref{fig:meas_circ} and \ref{fig:422code_meascirc2}. 
The last syndrome measurement in the protocol is performed ideally.
For the ancilla state that passes all of the post selections, 
we measure the logical $X$ operator and check whether the logical $Z$ error occurs. 
We estimate the failure rate of the post-selection and the logical $Z$ error rate of the prepared ancilla state 
by counting these events for all Monte Carlo samples. 
\begin{figure}[tbp]
  \centering
  \includegraphics[width=65mm, clip]{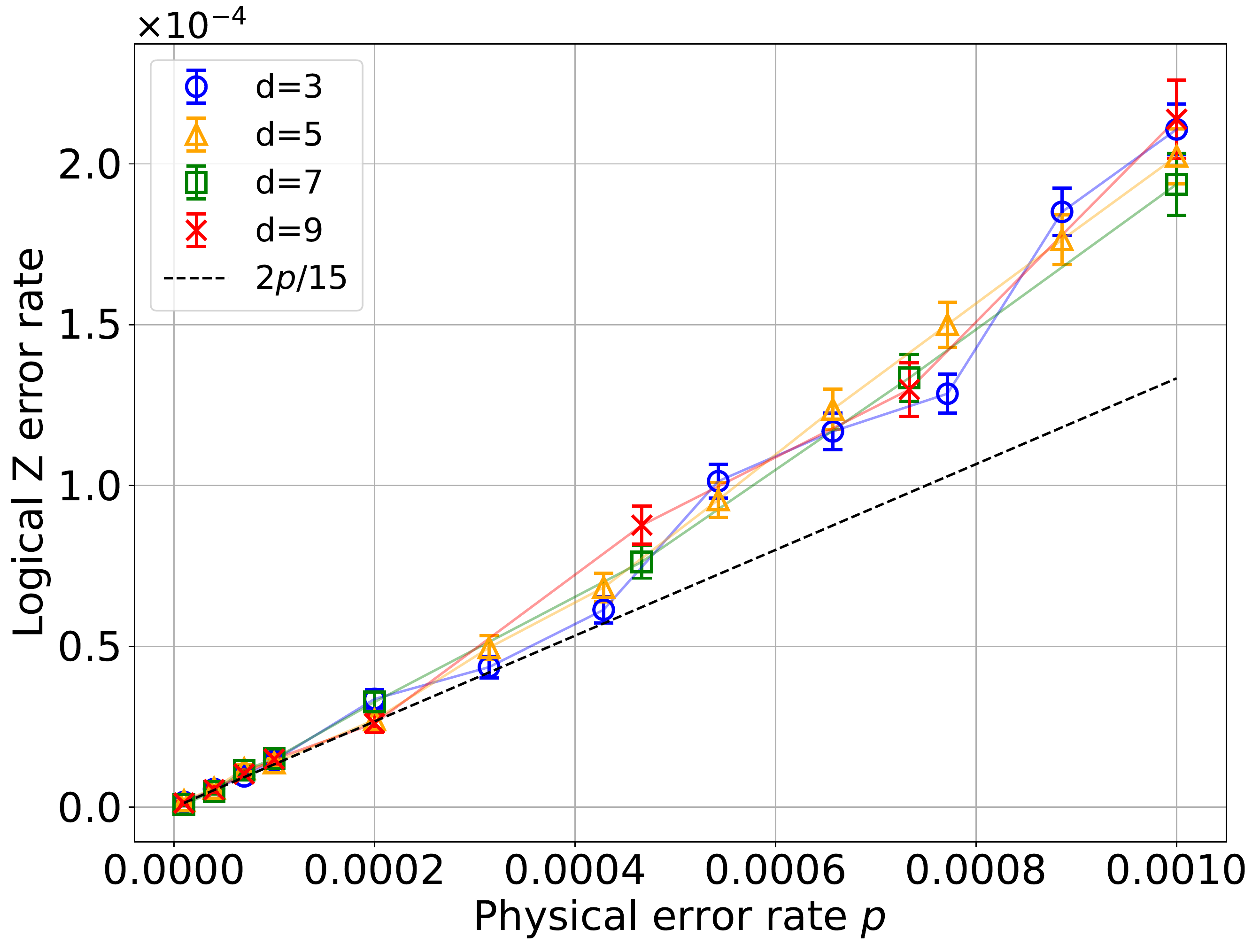} \\
  \includegraphics[width=65mm, clip]{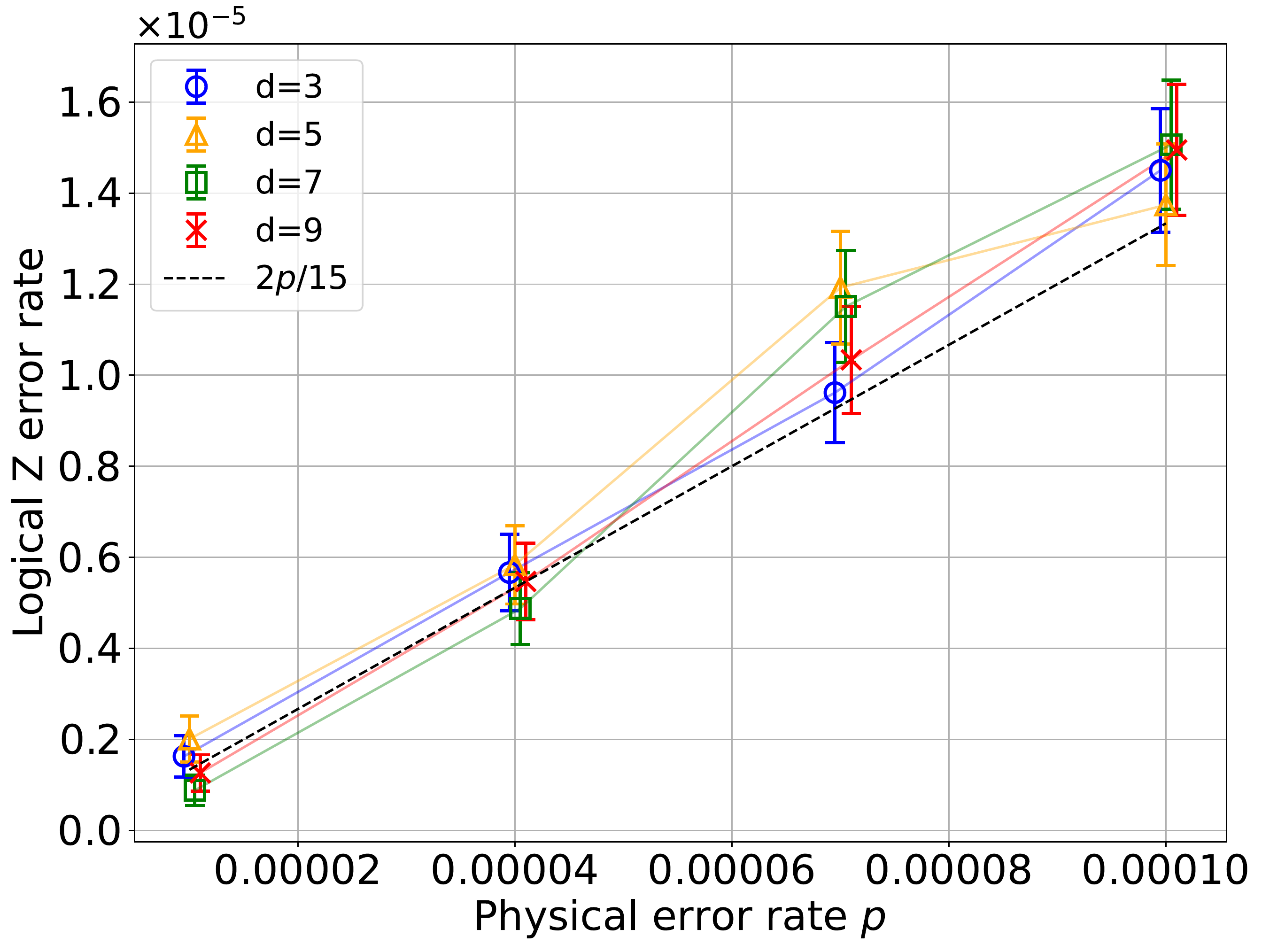}
  \caption{Logical $Z$ error rates of the ancilla state prepared in the surface code patches with $d = 3, 5, 7, 9$. 
  The dashed line shows the leading-order behavior expected under the circuit-level noise model, $P_{L, Z} (p) = 2p / 15$. 
  The error bars indicates $\pm 1 \sigma$ statistical errors estimated by the standard deviation. 
  (Upper) $p$ dependence in the range of $p \in [10^{-5}, 10^{-3}]$. 
  (Lower) Enlarged view in the range of $p \in [10^{-5}, 10^{-4}]$. }
  \label{fig:logerror_ancprep}  
\end{figure}
\begin{figure}[tbp]
  \centering
  \includegraphics[width=68mm, clip]{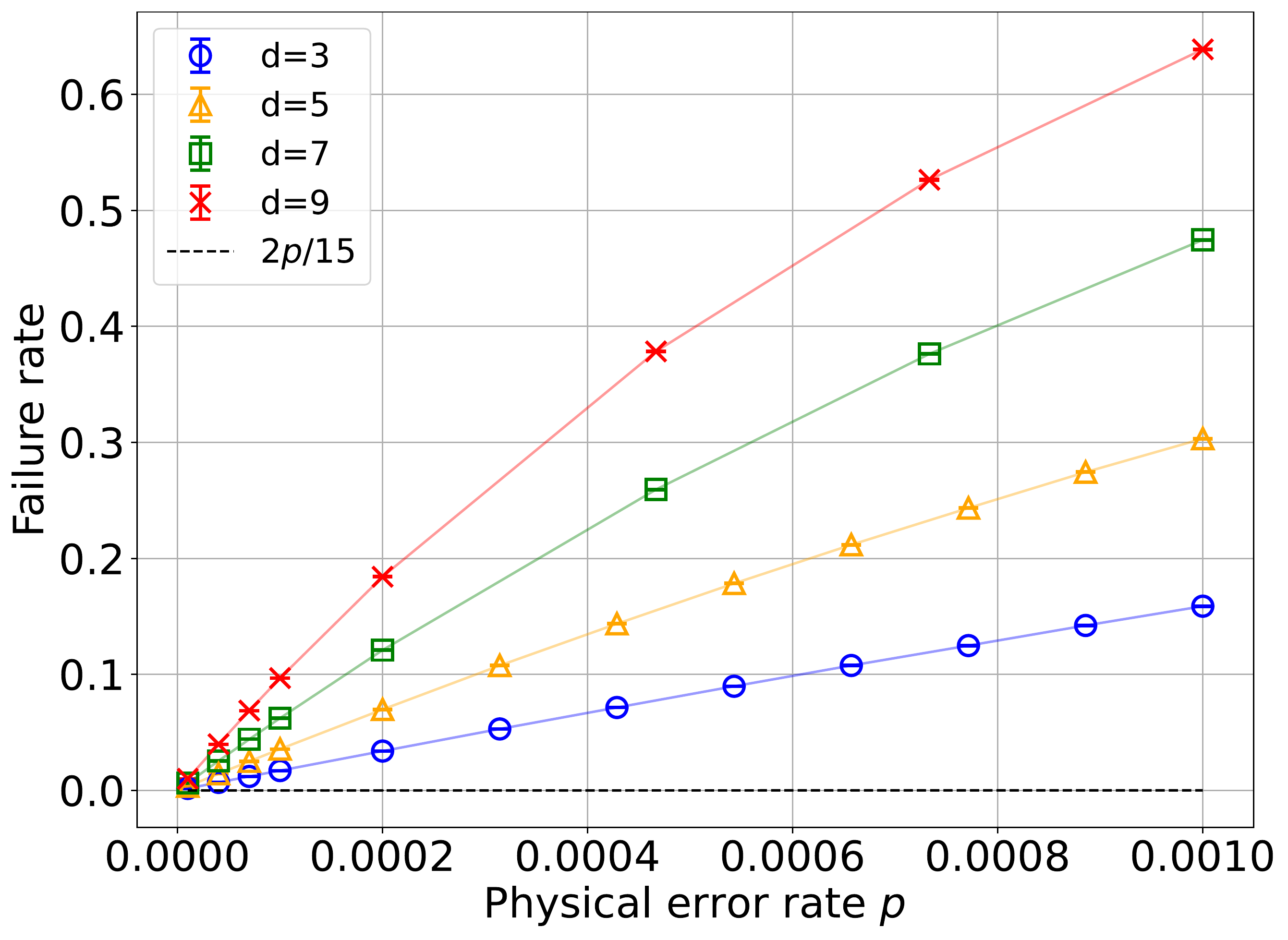} \\
  \includegraphics[width=68mm, clip]{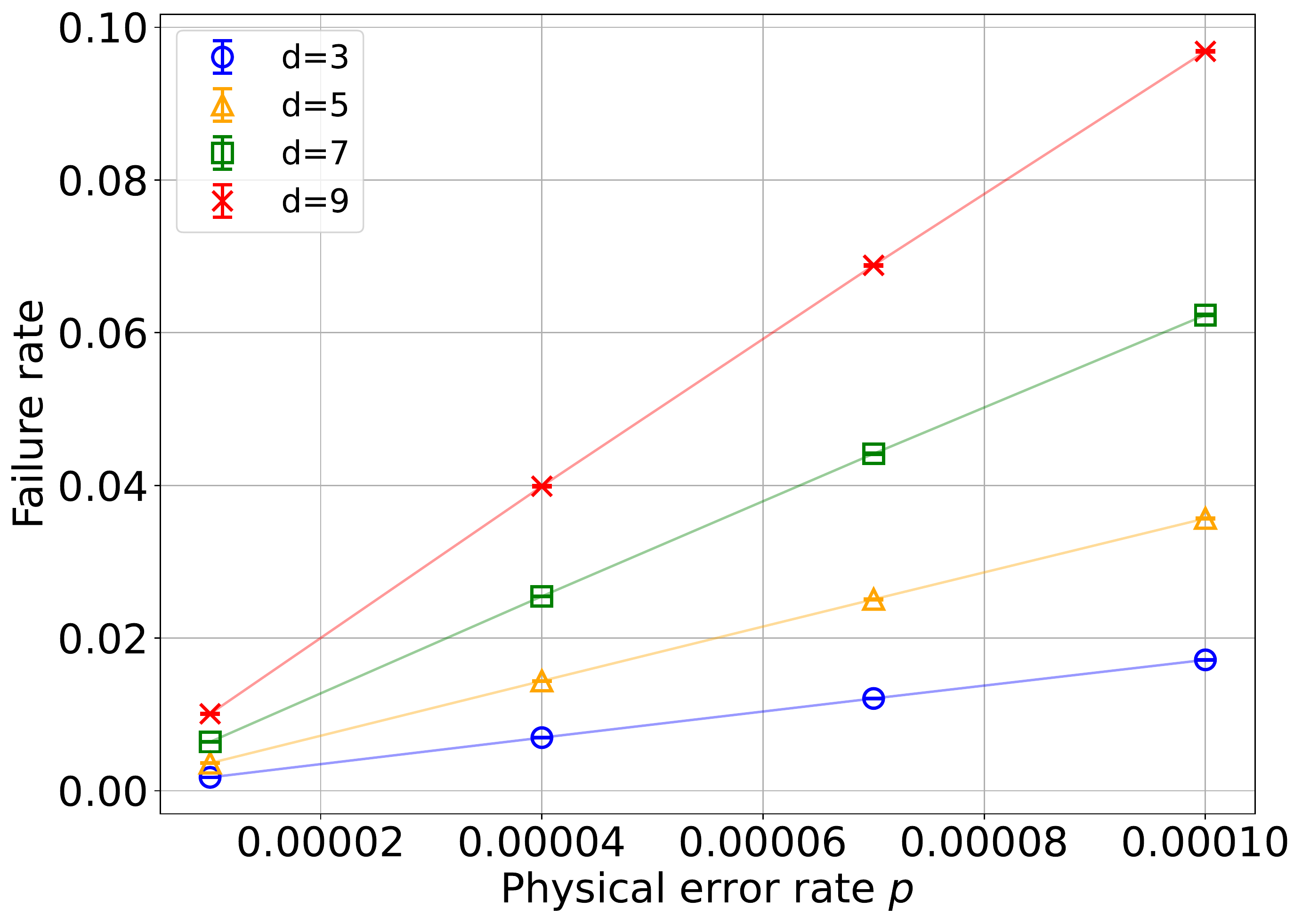}
  \caption{Failure probability of the post selection in our protocol. 
  (Upper) $p$ dependence in the range of $p \in [10^{-5}, 10^{-3}]$. 
  (Lower) Enlarged view in the range of $p \in [10^{-5}, 10^{-4}]$.}
  \label{fig:failure_ancprep}  
\end{figure}

For the resource estimation, we consider the target surface code patch with $d = 3, 5, 7,$ and $9$ 
and a physical error rate of $p \in [10^{-5}, 10^{-3}]$. 
The failure rate and the logical $Z$ error rate are estimated using $8 \times 10^6$ ($4 \times 10^6$) Monte Carlo samples 
with $p \in [10^{-5}, 10^{-4}]$ ($p \in [10^{-4}, 10^{-3}]$). 
Figures \ref{fig:logerror_ancprep} and \ref{fig:failure_ancprep} show the numerical results of the logical $Z$ error rate and the failure rate, respectively. 

First, we discuss the resultant logical $Z$ error rate. 
As discussed in Appendix~\ref{sect:appx1}, under the circuit-level noise model, the leading-order behavior of the logical $Z$ error rate is analytically given as $P_{L, Z} = 2p / 15 + O(p^2)$. 
Our numerical result in Fig.~\ref{fig:logerror_ancprep} (left) shows that the logical $Z$ error rate actually approaches the leading-order behavior when the physical error rate $p$ becomes small. 
Moreover, from Fig.~\ref{fig:logerror_ancprep} (right), we confirm that the sub-leading contribution of $O(p^2)$ can be neglected at a physical error rate below $10^{-4}$. 
Additionally, the logical $X$ error occurring in the aniclla state is negligible below $p = 10^{-4}$, although it is not directly confirmed. 
Next, we examine the failure rate of the post selection (Fig.~\ref{fig:failure_ancprep}). 
We observe that the failure rate increases when the code distance becomes longer. 
This behavior is due to the longer distance code having more possible error configurations at $O(p)$, which are captured by the second post-selection in our protocol. 
This large failure rate brings a large overhead for the state injection, and therefore we must reduce it using certain techniques. 
One solution is to repeat the protocol many times and reduce the effective failure rate. 
This can be achieved naively by the parallel injection using multiple patches, although this approach requires an additional space cost. 
Alternatively, we may reduce the effective failure rate without any additional space cost by parallelizing the protocol along the ``time direction".
To this end, one should first notice that the ancilla injection protocol can be performed within four rounds of the syndrome measurement (strictly, the total depth of the entire circuit can be $2 + 7 + 6 + 2 \times 8 = 31$ when we maximally overlap the circuits of Figs.~\ref{fig:meas_circ}, \ref{fig:422prep_circ}, and \ref{fig:422code_meascirc2}). 
If we consider the RUS protocol shown in Fig.~\ref{fig:scheme1_RUS}, 
we have a time interval of $2d$ rounds of the syndrome measurement during a single RUS step, 
and then we can repeat the state injection protocol roughly $2d / 4 = d/2$ times. 
With $p = 10^{-4}$ and $d = 9$, we have a failure rate of approximately 10\% as observed in Fig.~\ref{fig:failure_ancprep} (right), 
but $9/2 \sim 4$ repeats of the protocol effectively reduce the failure rate to 0.01\%. 

\subsection{Resource estimation} \label{sect:result_rsrc}
In this section, we estimate a computational resource for early-FTQC devices based on the results of the numerical simulations. 
Here we assume that a target device has $N=10^4$ physical qubits with a physical error probability of $p = 10^{-4}$. 

Initially, let us consider the number of logical qubits we can allocate. 
This number depends on the scheme for calculating a given circuit, as already discussed in Sec.~\ref{sec:lqa}. 
Since scheme (II) in Sec.~\ref{sec:lqa} is more efficient regarding the space cost, we consider it first. 
In the minimum construction using the compact block, we need at least $1.5n + 5$ logical patches to allocate $n$ data logical qubits. 
A single rotated surface code patch with the code distance $d$ needs $\approx 2d^2$ physical qubits, 
therefore $(1.5n + 5) \times 2d^2$ physical qubits are needed in total. 
If $N=10^4$, we can allocate $\approx 64$ ($\approx 37$) logical data qubits for the $d=7$ ($d = 9$) surface code patch in this setup. 
The same estimation for scheme (I) in Sec.~\ref{sec:lqa} brings 
$\approx 51$ ($\approx 30$) logical data qubits for $d=7$ ($d = 9$) if we employ the smallest arrangement of Fig.~\ref{fig:scheme1_arrange_alt} (b). 

Next, let us estimate the number of gate operations we can perform on those logical data qubits. 
Regarding the logical Clifford operation, 
by assuming that the error channel for the logical Clifford gate is an independent logical $Z$ and $X$ error channel, 
its logical error rate per $d$ rounds of the syndrome measurement can be given as $P_{L,round} \approx P_{L, Z} + P_{L, X}$, 
where $P_{L, Z}$ and $P_{L, X}$ are the logical error rates obtained by the numerical simulation discussed in Sec.~\ref{sect:result_surfacecode}.  
Using the fitting result of the logical error rates given in Tab.~\ref{tab:global_fit}, 
we can estimate $P_{L,round}$ as $P_{L,round} \approx 5.82 \times 10^{-8}$ ($P_{L,round} \approx 1.46 \times 10^{-9}$) for the $d = 7$ ($d = 9$) logical patch with $p = 10^{-4}$. 
The available number of the Clifford gates can be estimated as $N_{\rm Clifford} \approx 1 / P_{L,round}$, leading to $N_{\rm Clifford} \approx 1.72 \times 10^7$ ($N_{\rm Clifford} \approx 6.85 \times 10^8$) for the $d = 7$ ($d = 9$) logical patch with $p = 10^{-4}$. 
This number is sufficiently large, and $d = 7$ or $d = 9$ may be sufficient for most applications in the early-FTQC era. 
Note that in practice, the estimated $N_{\rm Clifford}$ may be divided by a certain $O(1)$ factor because some logical operations need 
more measurement rounds than $d$ rounds (e.g., the explicit logical CNOT needs $2d$ measurement rounds).
However, this factor is not expected to change the estimation drastically, 
so we only consider the estimated value of $N_{\rm Clifford}$ above as representative. 
The available number of analog rotation gates can be estimated similarly. 
For $p = 10^{-4}$, we can neglect the $O(p^2)$ contributions of the logical error rate as observed in the numerical simulation discussed in Sec.~\ref{sect:result_ancprep}. 
Therefore, a single rotation gate has a logical error rate of $P_{L, rotation} = 2p / 15 \approx 1.3 \times 10^{-5}$. 
Since the actual rotation gate needs two RUS steps on average, 
we can estimate the available number of the rotation gates as $N_{\rm rotation} \approx 1/(2P_{L, rotation}) = 3.75 \times 10^4$. 
As discussed in Sec.~\ref{sec:arg}, the remnant error of $N_{\rm rotation}$ analog rotations can be mitigated by the probabilistic error cancellation with an additional sampling overhead $\gamma^{2N_{\rm rotation}} \approx e^{8P_{L, rotation}N_{\rm rotation}} \approx 55$. 

In summary, assuming that $N = 10^4$ and $p = 10^{-4}$, the STAR architecture 
based on the $d = 7$ ($d = 9$) surface code patch can perform the quantum circuits using 64 (37) data qubits, 
which comprise $1.72 \times 10^7$ ($6.85 \times 10^8$) Clifford gates and $3.75 \times 10^4$ arbitrary rotation gates.
Notably, we can perform over $10^4$ arbitrary rotations and many error-corrected Clifford gates on $64$ logical qubits 
within a relatively lenient requirement, namely $N=10^4$ and $p = 10^{-4}$. 
Computations of this size cannot be simulated by classical supercomputers and state-of-the-art classical algorithms~\cite{PhysRevLett.116.250501,PRXQuantum.3.020361}. 
Even if we choose the $d = 9$ case, which classical supercomputers can simulate, 
it still provides a useful testbed for small-scale FTQC experiments through the direct comparisons with classical simulations. 

\subsection{Comparison to existing NISQ and FTQC architecture}
To clarify an advantage of our architecture, we compare its performance 
with those of naive NISQ architectures and existing FTQC architectures. 

\begin{figure}[tbp]
  \centering 
  \mbox{
    \Qcircuit @C=1.3em @R=1.0em {
      \lstick{\ket{0}} & \multigate{3}{\pi} & \multigate{1}{SU(4)} & \multigate{3}{\pi} & \multigate{1}{SU(4)} & \meter \\
      \lstick{\ket{0}} & \ghost{\pi}        & \ghost{SU(4)}        & \ghost{\pi}        & \ghost{SU(4)}        & \meter \\
      \lstick{\ket{0}} & \ghost{\pi}        & \multigate{1}{SU(4)} & \ghost{\pi}        & \multigate{1}{SU(4)} & \meter \\
      \lstick{\ket{0}} & \ghost{\pi}        & \ghost{SU(4)}        & \ghost{\pi}        & \ghost{SU(4)}        & \meter 
      \gategroup{1}{2}{4}{3}{.8em}{--}
      \gategroup{1}{4}{4}{5}{.8em}{--}
    }
  }
  \caption{
    Benchmark circuit used to measure the quantum volume with $m = 4$ and $d = 2$. 
    Operations grouped by dashed line form a single layer. 
  }
  \label{fig:quantum_volume_circuit}
\end{figure}
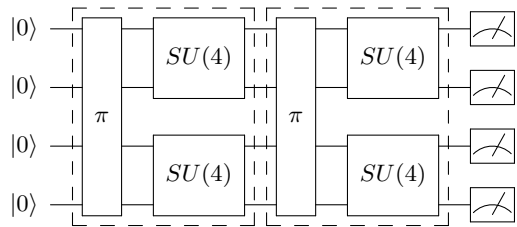

A typical performance metrics of the NISQ architecture is the quantum volume $V_Q$~\cite{PhysRevA.100.032328},
which quantifies the typical size of a correctly executable circuit. 
To measure $V_Q$, we utilize a benchmark circuit with $m$ qubits and $d$ layers, as shown in Fig.~\ref{fig:quantum_volume_circuit}. 
Each layer comprises a permutation (depicted by $\pi$ in Fig.~\ref{fig:quantum_volume_circuit}) and two-qubit unitary gates (depicted by SU(4) in Fig.~\ref{fig:quantum_volume_circuit}). 
The permutations and two-qubit unitaries in the benchmark circuit are randomly chosen and
an output distribution is determined by averaging over these randomly generated circuits. 
By considering the heavy output generation problem~\cite{PhysRevA.100.032328} on the output distribution, 
one can judge whether the circuit is implemented successfully: 
If the heavy output probability is more than two-thirds, the circuit is considered reasonably executed; otherwise the computation is failed. 
$V_Q$ is defined by the maximum size $m_{\rm max}$ of the square-shaped ($m = d$) benchmark circuit that is successfully implemented, 
$\log_2 V_Q = m_{\rm max}$.
To compare the performance between the STAR architecture and naive NISQ architecture, 
let us consider, for example, a quantum device comprising $10^4$ physical qubits with a square grid connectivity and an error rate of $p = 10^{-4}$. 
According to Ref.~\cite{PhysRevA.100.032328}, the size $m (= d)$ of the square-shaped circuit that can be executed correctly satisfies 
\begin{equation} \label{eq:qv_squaregrid}
  m^2 (1.29 \sqrt{m} - 0.78) p < 1,
\end{equation}
when the single-qubit error rate is negligible against the two-qubit error rate. 
The $\sqrt{m}$ factor comes from the restriction of the square grid connectivity. 
By inserting the physical error rate of $p = 10^{-4}$ into Eq.~(\ref{eq:qv_squaregrid}), we can estimate the quantum volume as $\log_2 V_Q = 37$
\footnote{
  This value may be overly large since it ignores single-qubit errors. 
  By adding the contribution of the single-qubit errors based on the transpiled SU(4) gate by Qiskit, we roughly obtain $m = 26$. 
  }. 
Regarding the STAR architecture, on the other hand, the permutations can be performed ideally and only the SU(4) gates suffer from errors. 
To quantify the error rate of the SU(4) gate, 
we use the fact that the SU(4) gate $U$ can be decomposed as $U = K_1 A(\alpha, \beta, \gamma) K_2$,
where $K_i (i=1,2)$ are tensor products of single-qubit unitary gates and 
$A(\alpha, \beta, \gamma) = \exp[i \left( \alpha X \otimes X + \beta Y \otimes Y + \gamma Z \otimes Z \right)]$~\cite{PhysRevA.100.032328}.
Each of the single-qubit unitary gates as well as $A(\alpha, \beta, \gamma)$ is implemented by three analog rotations.
Therefore, the entire SU(4) gate $U$ can be implemented by $3 \times 4 + 3 = 15$ analog rotations.  
Thus, the size $m$ of the square-shaped circuit that can be executed correctly satisfies 
\begin{equation}
  m^2 \times (15 / 2) \times 2.6 \times 10^{-5} < 1, 
\end{equation}
and an allowed maximum size is $m = 71$. 
Since the STAR architecture can prepare 64 logical qubits on $10^4$ physical qubits, it can achieve $\log_2 V_Q = 64$, which is substantially larger than that of the naive NISQ architecture. 
This result indicates that the STAR architecture reaches an advanced stage of quantum computation by successfully integrating error-corrected Clifford gates and noisy analog rotations when the physical error rate is sufficiently small. 
Note that this advantage gradually decreases if the physical error rate approaches the threshold value near $p_{\rm th} = 0.4$\% because the suppression of the logical error of the Clifford gates becomes poor. 
Therefore, to break through the difficulty of the NISQ architecture in the early-FTQC era, 
it is important to achieve a sufficiently small physical error rate below the threshold.

Next, we compare the STAR architecture with the existing FTQC architectures. 
Presently, the best space-efficient FTQC architecture is the one reported in Ref.~\cite{Litinski2019gameofsurfacecodes} and is introduced in Sec.~\ref{sect:latticesurgery}. 
To compare it to the STAR architecture, 
we first note that the STAR architecture can perform an analog multi-Pauli rotation 
with a logical error rate of $\epsilon = 2.6 \times 10^{-5}$ within 18 clocks on average 
(here, we call an execution time of $d$ rounds of the syndrome measurements as ``1 clock")
since the compact block consumes one magic state in 9 clocks in the worst case~\cite{Litinski2019gameofsurfacecodes}. 
In the following discussion, we estimate the resources needed to achieve the same performance of the analog multi-Pauli rotation using the existing FTQC architecture.  
On this basis, we estimate the available computational power of the existing FTQC architecture 
under the restriction of a quantum device that comprises $10^4$ physical qubits with a square grid connectivity and an error rate of $p = 10^{-4}$. 

The Clifford gates are implemented by the lattice surgery in both architectures; 
thus, we fix the code distance to $d = 7$ to make their performances even. 
Using the existing state injection protocol proposed in Ref.~\cite{Li_2015}, 
a bare magic state is obtained with an error rate $46p/15 \approx 3.1 \times 10^{-4}$. 
Because we must implement tens or hundreds of $T$ gates to perform analog rotation, 
this accuracy is insufficient and we must distill the magic state. 
Using the typical 15-to-1 distillation protocol~\cite{Litinski2019gameofsurfacecodes}, 
we obtain a clean magic state with an accuracy $35 \cdot (3.1 \times 10^{-4})^3 \approx 1.0 \times 10^{-9}$, 
whose precision is sufficient for our purpose. 
Therefore, to achieve an accuracy of $\epsilon = 2.6 \times 10^{-5}$ for a single analog rotation, 
the remaining task is to decompose the analog rotation gate into the sequence of Clifford gates and $T$ gates within the accuracy $\epsilon$. 
According to the state-of-the-art algorithm~\cite{ross2016optimal}, 
the required number of $T$ gates to achieve an approximation accuracy $\delta$ roughly behaves as $N \approx 3\cdot \log_2(1/\delta)$
\footnote{
  Precisely, authors in Ref.~\cite{ross2016optimal} conjectured that $N = K + 3\cdot \log_2(1/\delta)$ with some constant $K$ if $\delta \to 0$. 
  Our estimation may be rough since it neglects the constant term and $\delta = O(10^{-5})$ is not small. 
  However, similar algorithm proposed in Ref.~\cite{7056491} gives $N = 3.067\cdot \log_2(1/\delta) - 4.322$ around $\delta \in [10^{-2},10^{-10}]$, 
  therefore we expect that our estimation is reasonable up to $O(1)$ constant deviation. 
}. 
By substituting $\delta = \epsilon = 2.6 \times 10^{-5}$, we obtain $N \approx 46$. 
Thus, the existing FTQC architecture needs at least $46$ clocks to implement the analog rotation, 
which is approximately $2.6$-fold slower than the STAR architecture. 
It clearly show that the direct analog rotation is inherently advantageous for the fast operation. 

Furthermore, in the FTQC architecture, a trade-off relationship holds 
between the execution time of the non-Clifford gate and the number of the physical qubits 
because of the slow supply of the magic state by a single distillation block.
Let us consider the case in which a single $T$ gate takes 1 clock to implement. 
This case is related to the fast block in Ref.~\cite{Litinski2019gameofsurfacecodes}, 
which needs $2n+\sqrt{8n}+1$ patches to allocate $n$ logical qubits. 
In addition to the data block, we need a magic state factory that can supply one magic state per clock. 
According to Ref.~\cite{Litinski2019gameofsurfacecodes}, a single 15-to-1 distillation block can supply one magic state in 11 clocks using at least 11 patches. 
To achieve the required magic state supply rate, we must implement 11 distillation protocol in parallel; 
thus, at least $11 \times 11 = 121$ patches are required. 
If we consider the $d=7$ surface code, we can prepare at most $10^4 / (2\cdot 7^2) \approx 102$ patches 
and cannot even allocate the magic state factory. 
To reduce the physical qubit overhead, we can use other data blocks such as an intermediate block or compact block at the expense of operation speed. 
The intermediate block (compact block) takes 5 (9) clocks to implement a single $T$ gate~\cite{Litinski2019gameofsurfacecodes}, and the number of the distillation blocks in the factory can be reduced to 3 (2), respectively. 
The magic state factory requires $11 \times 3 = 33$ ($11 \times 2 = 22$) patches; 
thus, we can allocate logical qubits by using the remaining $102 - 33 = 69$ ($102 - 22 = 80$) patches. 
Because the intermediate block (compact block) requires $2n+4$ ($1.5n+3$) patches to allocate $n$ logical qubits, we can allocate $n=32$ ($n=51$) logical qubits. 
The intermediate block architecture can easily be simulated by existing classical supercomputers, and it is difficult to 
provide useful quantum advantages with the $10^4$ physical qubit device. 
Although the compact block architecture enters a classically intractable region, 
its available logical qubits are fewer than the STAR architecture ($n=64$) and its execution time is 23 times slower than ours. 
Therefore, the STAR architecture is advantageous in terms of the logical qubit number and execution time.  
We summarize these trade-off relationships of the existing FTQC architecture and their comparison to the STAR architecture in Table~\ref{tab:comparison_FTQC}. 
\begin{table*}
  \centering
  \caption{
    Comparison between the STAR architecture and existing FTQC architecture~\cite{Litinski2019gameofsurfacecodes} on the early-FTQC device 
    that consists of $10^4$ physical qubits with a square grid connectivity and an error rate $p = 10^{-4}$.
  }
  \begin{tabular}{cccc}
    \hline \hline
    Arch. & Num. of logical qubits & \quad Non-Clifford gate execution time [clock] \\ \hline
    STAR Compact ($d=7$) & 64 & 18 \\
    FTQC Fast ($d=7$) & 0 & 46 \\
    FTQC Intermediate ($d=7$) & 32 & 230 \\
    FTQC Compact ($d=7$) & 51 & 414 \\ \hline \hline
  \end{tabular}
  \label{tab:comparison_FTQC}
\end{table*} 

Although the STAR architecture always has advantages against the FTQC architecture in terms of the execution speed and the number of logical qubits, we should note that the executable number of the rotation gates is restricted by the inverse of the physical error rate. 
Therefore, the development of the low-error physical qubit is important again. 
In addition to hardware improvement, algorithmic improvement is also mandatory to achieve useful applications within the small number of rotation gates. 
If the fully-fledged FTQC becomes available in future, it will be necessary to use the STAR architecture and the FTQC architecture differently depending on the application. 
For example, quantum circuits in which the number of arbitrary rotations is not so large can be efficiently calculated using the STAR architecture. 
While in the case of quantum circuits that comprise an extremely large number of gate operations, the calculation is performed with high accuracy using the FTQC architecture. 

In summary, the STAR architecture can outperform a naive application of the NISQ and FTQC architecture to the early-FTQC device. 
Its advantage against the NISQ architecture is mainly attributed to the error-correction of the Clifford gates. 
Furthermore, compared to the existing FTQC architecture, we can say that 
the combination of the direct implementation of the analog rotation gate and the careful state injection protocol 
makes the STAR architecture faster and smaller with minimum compromising accuracy. 

\subsection{Possible applications}
Finally, we briefly discuss possible applications of the STAR architecture. 
Here, we only show some naive examples and typical calculation sizes based on the resource estimation. 
A detailed examination of the useful applications is an important future issue. 

One promising application of the STAR architecture is a quantum many-body simulation 
because the time-evolution operator can be implemented easily by analog rotation gates. 
For example, let us consider a 1D Hubbard model with $N$ sites. 
The Hamiltonian can be written in terms of Pauli operators as 
\begin{eqnarray}
  H = -t \sum_{i=0}^{2N-3} (X_i X_{i+2} + Y_i Y_{i+2}) Z_{i+1} \nonumber \\
  + \frac{U}{4} \sum_{i=0}^{N-1} Z_{2i-1} Z_{2i} - \frac{U}{4} \sum_{i=0}^{2N-1} Z_i, 
\end{eqnarray}
where $t$ and $U$ are parameters of the system and $P_i (P=X, Y, Z)$ are Pauli operators acting on the $i$-th degree of freedom. 
This Hamiltonian consists of $2N-2 + N + 2N = 5N -2$ terms; thus, its time evolution of a single Trotter step requires $5N-2$ arbitrary rotation gates. 
If we choose the $d=7$ ($d=9$) architecture and fully allocate logical data qubits for $N$ sites, we can simulate $N = 64 / 2 = 32$ ($N = 37 / 2 = 18$) sites. 
The actual number of rotation gates per Trotter step is $5\cdot 32-2 = 158$($5\cdot 18-2 = 88$). 
Therefore, we can simulate real-time dynamics with $3.75 \times 10^4 / 158 \approx 237$($3.75 \times 10^4 / 88 \approx 426$) Trotter steps for this system. 
More generally, for a Hamiltonian which have $64 \cdot (const)$ terms,  
the STAR architecture can simulate its real-time dynamics with $O(10^2)$ Trotter steps. 

Using the iterative phase estimation~\cite{PhysRevA.76.030306} or recent resource-efficient algorithms for the early-FTQC era~\cite{https://doi.org/10.48550/arxiv.2209.11322,https://doi.org/10.48550/arxiv.2211.11973}, 
phase estimation for unitary operators acting on $63$ or $37$ qubits can be achieved. 
This phase estimation can be applied to determining the ground state energy of the quantum system reachable in the STAR architecture 
if it allows sufficiently large Trotter steps. 
In this context, discretization errors must be minimized in the Trotterization. 
We may, for example, use the local variational quantum compiling (LVQC)~\cite{PRXQuantum.3.040302} for this purpose. 

Another promising application is the quantum approximation optimization algorithm (QAOA)~\cite{https://doi.org/10.48550/arxiv.1411.4028} for solving binary optimization problems. 
For example, let us consider the MaxCut problem of a graph with $N$ nodes. 
The problem Hamiltonian is given as 
\begin{eqnarray}
  H_C = -\frac{1}{2} \sum_{i \neq j} ( 1- Z_i Z_j ).
\end{eqnarray}
To obtain the ground state of $H_C$, 
we consider the QAOA ansatz state, 
\begin{eqnarray}
  \ket {\gamma, \beta} = e^{-i\beta_{p-1} H_B} e^{-i \gamma_{p-1} H_C} \cdots e^{-i\beta_{0} H_B} e^{-i \gamma_{0} H_C} \nonumber \\ 
  \times H^{\otimes N} \ket{0}^{\otimes N},
\end{eqnarray}
where 
\begin{equation}
  H_B = \sum_{j=0}^{N-1} X_j, 
\end{equation} 
and ${\bf \gamma} = (\gamma_0, \cdots \gamma_{p-1})$, ${\bf \beta} = (\beta_0, \cdots \beta_{p-1})$ are optimization parameters. 
They are optimized to minimize an expectation value $\bra {\gamma, \beta} H_C \ket {\gamma, \beta}$. 
The ansatz state contains $p\cdot (N + \frac{N(N-1)}{2})$ arbitrary rotations in total. 
If we choose the $d=7$ ($d=9$) architecture and set $N=64$ ($N=37$), 
we can take the depth of the ansatz as $p = 3.75 \times 10^4 / 2080 \approx 18 (3.75 \times 10^4 / 703 \approx 53)$. 
Note that higher order binary optimization (HOBO) problems can be directly solved in the STAR architecture without any reduction to quadratic unconstrained binary optimization (QUBO) problems, because the Clifford gates are almost error-free. 

\section{Conclusion} \label{sec:conclusion}
In this work, we propose a quantum computing architecture suitable for the early-FTQC devices, the STAR architecture. 
In the STAR architecture, 
universal quantum computation is achieved by arbitrary rotation gates and error-corrected Clifford gates. 
Analog rotation gates are realized by the RUS protocol with appropriate ancilla states. 
To reduce logical errors of the rotation gates, we carefully design the ancilla state injection protocol 
by combining the $[[4,1,1,2]]$ subsystem code and post-selection. 
Thus, our rotation gate achieves a small logical error rate of $P_L = 2p / 15 + O(p^2)$ under the circuit-level noise model, 
which is verified numerically. 
Clifford operations are performed by the standard lattice surgery protocol based on the rotated surface code, 
and we illustrate typical logical qubit arrangements. 
We also perform a numerical simulation on the surface code patch and determine a scaling behavior of the logical error rate. 
Finally, we estimate an available computational resource in the STAR architecture under the assumption of typical early-FTQC devices, 
where $N=10^4$ physical qubits can operate with a gate fidelity of $p = 10^{-4}$. 
According to this estimate, we can act $3.75 \times 10^4$ arbitrary rotation gates and 
$1.72 \times 10^7$ Clifford gates on 64 logical qubits encoded in the $d = 7$ rotated planar surface code. 
Classical computers cannot emulate such computations. 
Furthermore, the STAR architecture can surpass the naive NISQ architecture and the existing FTQC architecture. 
The STAR architecture may apply to some useful applications such as quantum many-body simulation, phase estimation, and QAOA. 

Some topics are not addressed in this paper. 
Here, we summarize these topics to envision the future directions of our proposal. 
(i) The optimization of the logical qubit arrangement and input quantum circuit. 
Regarding the logical qubit arrangement, we only illustrate some prototypical arrangements in this paper. 
In practical applications, however, the number of logical operations that can be performed simultaneously must be maximized to reduce computational time. 
Such parallelization highly depends on the structure of the quantum circuit we want to perform. 
Therefore, we must develop a clever compiler that decomposes the input circuit into the sequence of Clifford gates and $R_Z(\theta)$, and 
determines the patch arrangement maximizing the gate parallelism based on the decomposed circuit. 
(ii) More concrete discussion on the possible applications of the STAR architecture. 
In this study, we only briefly mentioned some prototypical quantum computations that can be performed on the STAR architecture. 
By combining clever resource-reduction techniques such as LVQC, 
the STAR architecture may give us some useful applications at the earlier stage of a large-scale quantum device. 
(iii) Improvements in our injection protocol. 
Although our injection protocol minimizes the remaining logical error on the ancilla state, it still lives on $O(p)$. 
To perform more interesting computations, we must further reduce the logical error rate on the analog rotation. 
However, the distillation protocol on the arbitrary rotation ancilla state has not been known until now, 
and the task of reducing its logical error rate to $O(p^2)$ is challenging. 
Developing a more sophisticated state injection/distillation protocol for early-FTQC era is an interesting future direction. 

We hope that our proposal and the corresponding development of quantum
algorithms will bring new insights to realizing practical quantum computers in future.

\begin{acknowledgments}
We would like to thank Jun Fujisaki and Mitsuki Katsuda for fruitful discussions. 
K.F. is supported by MEXT Quantum Leap Flagship Program (MEXT Q-LEAP) Grant No. JPMXS0118067394 and JPMXS0120319794, 
JST COI-NEXT Grant No. JPMJPF2014, and JST Moonshot R\&D Grant No. JPMJMS2061.
\end{acknowledgments}

\appendix
\section{Leading-order logical error probability of $\ket{m_{\theta}}_L$ under the circuit-level noise model} \label{sect:appx1}
In this appendix, we discuss a leading-order logical error probability of the ancilla state prepared in our protocol under the circuit-level noise model. 
We consider the same noise model discussed in Sec.~\ref{sect:result_surfacecode}: All physical operations suffer from error, which occur with a common probability $p$.

First, we consider logical errors occurring in the ancilla state injection circuit of Fig.~\ref{fig:422prep_circ}. 
In this circuit, there can be weight-2 logical errors with the probability proportional to $p$, 
due to the error propagation of the CNOT operations and the two-qubit depolarizing channels. 
The error propagation of the CNOT operations causes weight-2 errors, such as 
\begin{equation} \label{eq:error_prep1}
  X_0 X_1, X_2 X_3, Z_0 Z_1, Z_2 Z_3.
\end{equation}
Note that errors on qubits 2 and 3 are identical to those on qubits 0 and 1 up to stabilizer operators. 
Since $Z_0 Z_1 (Z_2 Z_3)$ is the logical $Z$ error on the gauge DOF, it is not critical for state injection. 
The other one, $X_0 X_1 (X_2 X_3)$, is the logical $X$ error on the logical qubit, 
but it does not destruct the logical state since the state is ${\ket +}_L$ at that moment. 
The single $Y$ errors before CNOT operation also lead another weight-2 errors, e.g., $Y_0 \otimes Z_1$ or $X_0 \otimes Y_1$, 
but those are detected as single $X_0$ or $Z_1$ errors and removed by the post-selection. 
In the same discussion, weight-2 errors produced by the two-qubit depolarizing channel in the noisy CNOT operation do not destruct the logical state. 
Regarding the noisy $R_{Z_0 Z_2}(\theta)$, however, there are weight-2 errors that destruct the logical state. 
The two-qubit depolarizing channel after the ideal $R_{Z_0 Z_2} (\theta)$ provides weight-two errors, such as
\begin{equation} \label{eq:error_prep2}
  Z_0 Z_2, X_0 X_2.
\end{equation}
In those examples, $X_0 X_2$ is the logical $X$ error on the gauge DOF and does not affect the logical state. 
On the other hand, $Z_0 Z_2$ is the logical $Z$ error and changes the ancilla state to an orthogonal state as follows,
\begin{eqnarray}
  Z_L \ket{m_\theta}_L =  Z_L \left( e^{-i\theta/2} {\ket 0}_L + e^{i\theta/2} {\ket 1}_L \right) \nonumber \\
  = e^{-i\theta/2} {\ket 0}_L - e^{i\theta/2} {\ket 1}_L \equiv {\ket {\overline m_\theta}}_L. 
\end{eqnarray}
Weight-2 errors that cause the logical $Z$ error in the two-qubit depolarizing channel are $Z_0 Z_2$ and $Y_0 Y_2$. 
Therefore its occurring probability is $2p / 15$. 
There are other weight-2 errors like $Y_0 X_2$, 
but they are detectable as a single qubit error. 
In addition, there are some $O(p)$ error propagation processes that cause an inverse rotation (= logical $X$ error), 
such as $R_{Z_0 Z_2}(\theta) X_0 = X_0 R_{Z_0 Z_2}(-2\theta) \cdot R_{Z_0 Z_2}(\theta)$, 
but those errors are detectable as a single qubit error. 
In summary, the logical error rate of the ancilla state generated by the circuit of Fig.~\ref{fig:422prep_circ} behaves as 
\begin{eqnarray}
  P_{Z_L}(p) &=& 2p / 15 + O(p^2), \label{eq:422prepcirc_logzerr} \\
  P_{X_L}(p) &=& O(p^2). \label{eq:422prepcirc_logxerr}
\end{eqnarray}

\begin{figure}[h]
  \centering
  \mbox{
    \Qcircuit @C=1em @R=.6em @!R {
        \lstick{M1} & \push{\ket{0}} \qw & \targ     & \push{\textcolor{blue}{Z}} \qw & \targ     & \meter \\
        \lstick{0}  & \qw                & \ctrl{-1} & \push{\textcolor{blue}{Z}} \qw & \qw       & \push{\textcolor{blue}{Z}} \qw \\
        \lstick{1}  & \qw                & \qw       & \qw                            & \ctrl{-2} & \push{\textcolor{blue}{Z}} \qw 
        \gategroup{1}{4}{2}{4}{.7em}{.}     
    }
  }
  \caption{Example of a $O(p)$ weight-2 error in the measurement circuit of the $[[4,1,1,2]]$ subsystem code.
  The $Z \otimes Z$ error occurring the first CNOT gate (grouped by a dotted line) propagates to physical qubits and forms a weight-2 error.
  }
  \label{fig:422code_meascirc_error}
\end{figure}
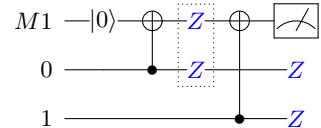

Next, we examine possible logical errors during the syndrome measurement circuits of Figs.~\ref{fig:meas_circ} and \ref{fig:422code_meascirc2}. 
In the circuit of Fig.~\ref{fig:422code_meascirc2}, a single error on the measurement qubit leads at most weight-1 errors on physical qubits, and they do not lead to undetectable logical errors.  
One possibility to realize $O(p)$ weight-2 errors is the two-qubit error occurring in the first CNOT operation with error propagation through the second CNOT operations, as shown in  Fig.~\ref{fig:422code_meascirc_error}. 
However, this weight-2 error acting on physical qubits is the gauge operator and is absorbed by the gauge DOF. 
Thus, the measurement circuit of Fig.~\ref{fig:422code_meascirc2} does not amplify the leading-order logical error rate of the ancilla state, Eqs.(\ref{eq:422prepcirc_logzerr}) and (\ref{eq:422prepcirc_logxerr}). 
Regarding the syndrome measurement circuit of Fig.~\ref{fig:meas_circ}, 
weight-2 errors can occur but they are orthogonal to the logical operators due to the order of CNOT operation. There is no other possibility to generate logical errors at $O(p)$. 
Therefore, the circuit of Fig.~\ref{fig:meas_circ} does not amplify the leading-order behavior of the logical error rate as well. 

In conclusion, the logical error rate of the ancilla state $\ket{m_\theta}$ prepared in our protocol is dominated by the ancilla state injection circuit (Fig.~\ref{fig:422prep_circ}) and behaves as $P_L = 2p / 15 + O(p^2)$.

\bibliographystyle{apsrev4-1}
\bibliography{bib.bib}

\end{document}